\documentclass[preprint,12pt]{elsarticle}
\usepackage[margin=0.8in]{geometry}  % 设置页边距为1英寸
\usepackage{amsmath}  
\usepackage{amssymb}
\usepackage{multirow}
\usepackage{array}
\usepackage{caption}
\usepackage{natbib}
\usepackage{xcolor}
\usepackage{colortbl}
\usepackage{xurl}
\usepackage{hyperref}
\usepackage{rotating} 
\usepackage{booktabs}
\usepackage{subcaption}
\usepackage{eurosym}

\newcommand{\revblue}[1]{\textcolor{black}{#1}}

\usepackage[linesnumbered,ruled,vlined]{algorithm2e}
\journal{}
\begin{document}

\begin{frontmatter}

\title{A Hierarchical Optimisation Framework for Integrated Electric-Hydrogen-Transport Systems}

\author[inst1]{Fulong Yao\corref{cor1}}
\ead{yaof@cardiff.ac.uk}
\affiliation[inst1]{organisation={School of Computer Science and Informatics},
            addressline={Cardiff University}, 
            city={Cardiff},
            postcode={CF10 3AT}, 
            % state={State One},
            country={UK}}
\author[inst2]{Yiming Xu}
\ead{XuY130@cardiff.ac.uk}
\affiliation[inst2]{organisation={School of Engineering},%Department and Organization
            addressline={Cardiff University}, 
            city={Cardiff},
            postcode={CF24 3AA}, 
            % state={State Two},
            country={UK}}
            
\author[inst2]{Liana Cipcigan}
\ead{CipciganLM@cardiff.ac.uk}
\author[inst2]{Maurizio Albano}
\ead{AlbanoM@cardiff.ac.uk}
\author[inst1]{Naeima Hamed}
\ead{HamedNH@cardiff.ac.uk}
\author[inst1]{Nima Valizadeh}
\ead{ValizadehN@cardiff.ac.uk}
\author[inst1]{Omer Rana}
\ead{RanaOF@cardiff.ac.uk}

\cortext[cor1]{Corresponding author: Fulong Yao} 

% \author{Anonymous authors}

\begin{abstract}

Integrated electric–hydrogen infrastructures are becoming increasingly important with the growing deployment of electric vehicles (EVs) and hydrogen vehicles (HVs) in transport systems. However, the strong coupling between vehicle scheduling and multi-energy dispatch introduces significant operational challenges. This paper models an integrated electric–hydrogen–transport system (EHTS) and proposes a hierarchical optimisation framework that couples vehicle scheduling and downstream energy dispatch through a sequential, demand-driven two-layer structure. In the vehicle scheduling layer, a solver-free greedy heuristic (SFGH) algorithm is developed to avoid repeated optimisation solving, enabling real-time EV charging and HV refuelling under non-preemptive service and within-interval sequential assignment. The resulting charging and refuelling demands are subsequently passed to the energy dispatch layer, where a deep reinforcement learning (DRL)-based approach is designed to optimise battery operation, hydrogen-tank operation, and PV generation allocation to promptly satisfy the scheduled transport demand while reducing the overall operational cost of the EHTS. Representative case studies, together with comparative, ablation, and generalisation analyses, demonstrate the effectiveness and robustness of the proposed framework. Compared with the conventional first-in-first-out (FIFO) scheduling algorithm, the proposed SFGH reduces the test-average cumulative EV/HV service delay by 206 h. Among all evaluated DRL approaches, Twin Delayed Deep Deterministic Policy Gradient (TD3) achieves the lowest overall operational costs, reducing annual costs by approximately £157k relative to the no-energy storage system (No-ESS) baseline and by about £52k compared with the rule-based approach. Furthermore, the learned dispatch policy maintains strong performance across diverse transport-demand scenarios and PV penetration levels without retraining, demonstrating robust generalisation capability for practical deployment.

\end{abstract}

\end{frontmatter}

%% main text
\section{INTRODUCTION} \label{sec:1}
The rapid decarbonisation of road transportation is driving the deployment of multiple low-carbon vehicle technologies, particularly electric vehicles (EVs) and hydrogen vehicles (HVs) \cite{iea2025ev}.
While EVs are widely regarded as an effective solution for passenger transport, HVs provide complementary advantages for heavy-duty and long-distance applications owing to their fast refuelling capability and high energy density \cite{cpcatapult2025hgv}.
Consequently, future transport systems are expected to support the long-term coexistence of EVs and HVs, driving the development of integrated charging–refuelling hubs rather than separate infrastructures.
This trend is already reflected by recent industrial developments. The UK company Aegis Energy, for instance, has secured a £100 million investment to accelerate the deployment of a nationwide network of clean multi-energy hubs for commercial vehicles, offering integrated electric charging and hydrogen refuelling within a single station \cite{reuters_aegis2025}. Such electric–hydrogen–transport systems (EHTSs) tightly integrate transport services with coupled electricity–hydrogen energy infrastructures, where heterogeneous vehicle demand interacts directly with energy supply, conversion, and storage within a unified operational system \cite{chen2025superconducting}.

Operating such systems is particularly challenging because both transport demand and energy resources are highly dynamic. Stochastic vehicle arrivals continuously change charging and refuelling demand, while renewable generation, electricity prices, and electricity–hydrogen resources also vary over time \cite{wan2025coordinated}. Efficient operation relies on two coupled decision processes: vehicle scheduling determines when and how arriving EVs and HVs are served, thereby setting the temporal charging and refuelling demand \cite{wan2025coordinated}, whereas energy dispatch allocates renewable generation, grid electricity, and hydrogen resources to satisfy that demand economically \cite{fang2023optimal}. Capturing this scheduling-to-dispatch relationship is therefore central to efficient EHTS operation. Accordingly, the following review examines existing studies from three perspectives: vehicle-service scheduling, electricity–hydrogen energy management, and integrated EHTS operation, in order to clarify how the demand observed by the dispatch layer is currently represented and where the  scheduling-to-dispatch dependency remains insufficiently characterised.

Research on vehicle-service scheduling has primarily focused on improving the utilisation of charging/refuelling facilities and reducing vehicle waiting time through real-time scheduling and resource allocation \cite{salam2024charge,zhou2024comprehensive}. Owing to the rapid deployment of battery-electric vehicles and charging infrastructures, EV charging scheduling has received sustained research attention over the past decade. For instance, 
\cite{zhao2024two} proposed a two-layer EV scheduling framework integrating online reservation and dynamic pricing to coordinate heterogeneous charging requests, thereby alleviating charging congestion and flattening aggregate grid load. \cite{tsaousoglou2023fair} developed a fair EV charging scheduler under distribution-network constraints by explicitly balancing charging delay and user fairness while respecting network operating limits. In contrast, research on hydrogen vehicles has remained comparatively limited because hydrogen-refuelling infrastructures are still at an earlier stage of deployment. Existing studies have mainly focused on infrastructure planning and queue analysis rather than real-time scheduling. For example,
\cite{zheng2014queuing} proposed a queuing-based optimisation framework to determine the minimum number of hydrogen dispensers required to satisfy prescribed waiting-time and queue-length constraints, whereas \cite{brown2022analysis} analysed large-scale operational data from retail hydrogen stations to characterise customer queuing behaviour and quantify the impact of dispenser capacity on service waiting times. Although these studies have improved the operational efficiency of individual EV charging or HV refuelling systems, they optimise the two services independently, without jointly scheduling heterogeneous EVs and HVs within an integrated station.

Recently, several studies have begun to jointly consider EVs and HVs within integrated transport systems. For example, \cite{wan2025coordinated} proposed a Stackelberg--Nash game framework that jointly models the spatiotemporal dynamics and charging/refuelling preferences of EVs and HVs to optimise microgrid operation, while  \cite{li2024cooperative} developed a cooperative scheduling framework for an electric--hydrogen integrated energy system considering coordinated EV/HV participation through V2G response and carbon trading. These studies represent important advances towards integrated EV/HV operation. However, they primarily coordinate vehicle participation at the transportation-network or integrated-energy-system level, rather than explicitly scheduling individual heterogeneous vehicles within an integrated charging--refuelling station. 
% More importantly, the charging and refuelling demand generated by vehicle scheduling is not explicitly represented as the endogenous input to downstream electricity--hydrogen energy dispatch. 
Consequently, a station-level scheduling framework capable of jointly accommodating heterogeneous EV and HV services while dynamically generating charging and refuelling demand for downstream multi-energy management remains insufficiently investigated.

Another body of research addresses how energy demand is met through integrated electricity–hydrogen system management. Conventional optimisation approaches, including mixed-integer linear programming (MILP), model predictive control (MPC), genetic algorithm (GA), and stochastic optimisation, have been widely adopted to coordinate renewable generation, battery energy storage, electrolysers, and hydrogen storage under various operational objectives \cite{tan2026distributionally,li2026optimal,elsir2025holistic,sun2025end}. However, these optimisation-based approaches typically require repeatedly solving mathematical optimisation problems whenever system states or uncertainties change. As the scale and complexity of coupled electricity–hydrogen systems increase, such repeated optimisation can become computationally demanding, thereby limiting their applicability to fast real-time operation. Deep reinforcement learning (DRL) has recently emerged as a promising alternative because it learns control policies directly through interactions with the environment and enables fast online decision-making once training is completed.
For instance, authors in \cite{yao2025unified,yao2025holistic} investigated value-based DRL algorithms, including Deep Q-Network (DQN), Double DQN (DDQN), and Double Duelling DQN (D3QN), for the coordinated control of battery and hydrogen storage in integrated electricity–hydrogen energy systems to reduce operational costs and flatten aggregate load profile.  Reference \cite{jadidbonab2026drl} employed two actor–critic DRL algorithms, namely Proximal Policy Optimisation (PPO) and Twin Delayed Deep Deterministic Policy Gradient (TD3), to perform energy dispatch in an integrated electricity–hydrogen hub under renewable-generation and demand uncertainties, with the objective of minimising long-term operational costs. Work in \cite{zhang2025hydrogen} employed Soft Actor–Critic (SAC) and Deep Deterministic Policy Gradient (DDPG) to coordinate energy management in hydrogen-integrated power systems for voltage regulation and network-loss reduction. Despite these advances, such studies generally treat the electricity and hydrogen demands as predefined, forecasted, or externally specified inputs. This assumption, however, does not hold once energy management is coupled with transport services: the charging and refuelling demand is not an externally given quantity but is itself produced by how individual vehicles are scheduled and served at the station. 

A third and more recent body of research has begun to investigate integrated EHTSs by jointly considering transport systems and downstream electricity–hydrogen energy management. Beyond the system-level coordination of \cite{wan2025coordinated} and \cite{li2024cooperative} discussed above, \cite{zheng2025safe} developed a distributionally robust optimisation framework for an electric--hydrogen refuelling station considering vehicle-arrival uncertainty, hydrogen-storage safety, and station profitability, while \cite{abdelghany2026optimal} jointly optimised hydrogen refuelling schedules with hydrogen production, storage, and electricity-market participation under renewable and price uncertainties. These studies have substantially advanced the integration of transport systems with coupled electricity--hydrogen energy infrastructures from both system-level and station-level perspectives. Nevertheless, transport demand is generally represented through aggregate traffic behaviour \cite{wan2025coordinated}, system-level vehicle participation \cite{li2024cooperative}, given demand profiles \cite{zheng2025safe}, or forecasted charging/refuelling demand \cite{abdelghany2026optimal}, rather than being generated through real-time scheduling. As a result, downstream energy dispatch is optimised against assumed demand profiles rather than dynamically evolving transport demand.

Overall, although recent studies have significantly advanced vehicle-service scheduling, electricity–hydrogen energy management, and integrated EHTS operation, a station-level framework that jointly schedules heterogeneous EVs and HVs and, in doing so, generates the charging and refuelling demand that drives downstream multi-energy dispatch is still lacking. Across the three bodies reviewed above, transport demand is generally treated as an aggregated or predefined input to the energy system rather than as the endogenous product of real-time heterogeneous vehicle scheduling. 
Therefore, dispatch is optimised in isolation from the scheduling decisions that actually produce its demand, leaving one dependency unaddressed: the operational link between heterogeneous EV/HV scheduling and electricity–hydrogen management.

To address this gap, this paper proposes a sequential hierarchical optimisation framework for EHTSs. The framework consists of a vehicle-scheduling layer and an energy-dispatch layer sequentially coupled through dynamically generated charging and refuelling demand. Specifically, the vehicle-scheduling layer minimises service delay by scheduling heterogeneous EV and HV charging/refuelling requests, while the energy-dispatch layer dispatches electricity and hydrogen resources to satisfy the resulting energy demand at minimum operational costs involving PV curtailment,  electricity purchases, and hydrogen procurement. By explicitly bridging transport-service scheduling and downstream multi-energy dispatch through demand propagation, the proposed framework enables demand-responsive operation of coupled electricity and hydrogen infrastructures under dynamically varying transport demand. The main contributions of this paper are summarised as follows:

1) A hierarchical optimisation framework is proposed for integrated EHTSs, in which transport-service scheduling and multi-energy dispatch are organised into two sequentially coupled layers connected through dynamically inferred charging and refuelling demand, making explicit the scheduling-to-dispatch dependency that prior work leaves unmodelled.

2)  An integrated operational model is developed for EHTSs that incorporates transport-service scheduling, charging/refuelling facilities, coupled electricity–hydrogen infrastructures, renewable generation, and battery/hydrogen storage into a unified operational model. In particular, the model introduces a non-preemptive, within-interval sequential assignment scheme that captures realistic facility-sharing among heterogeneous EVs and HVs, so that the charging and refuelling demand emerges from the scheduling process itself rather than being specified in advance.

3) A solver-free greedy heuristic (SFGH) is developed to solve the vehicle-scheduling problem without repeated online optimisation, and a TD3-based dispatch strategy is developed to coordinate battery, hydrogen-tank, and PV operation. The dispatch strategy adopts a benefit-based reward and a history-based state representation tailored to the heterogeneous cost scales and high uncertainty of electricity–hydrogen operation. Together they realise the framework for real-time operation, enabling efficient online scheduling and cost-effective multi-energy management.

4) Comprehensive comparative, ablation, and generalisation studies are performed using hybrid simulation datasets to evaluate the proposed framework against state-of-the-art scheduling and energy-dispatch methods, while further validating its robustness under diverse transport-demand scenarios and PV Penetration levels.

The remainder of this paper is organised as follows. Section \ref{sec:2} presents the integrated electric–hydrogen-transport system model. Section \ref{sec:3} introduces the proposed hierarchical optimisation framework. Section \ref{sec:4} presents  simulation setup and result analysis. Section \ref{sec:5} concludes the paper.

\section{EHTS Modelling} \label{sec:2}
 \begin{figure*}[]
	\begin{center}
        \includegraphics[width=12cm, trim=5 5 5 5, clip]{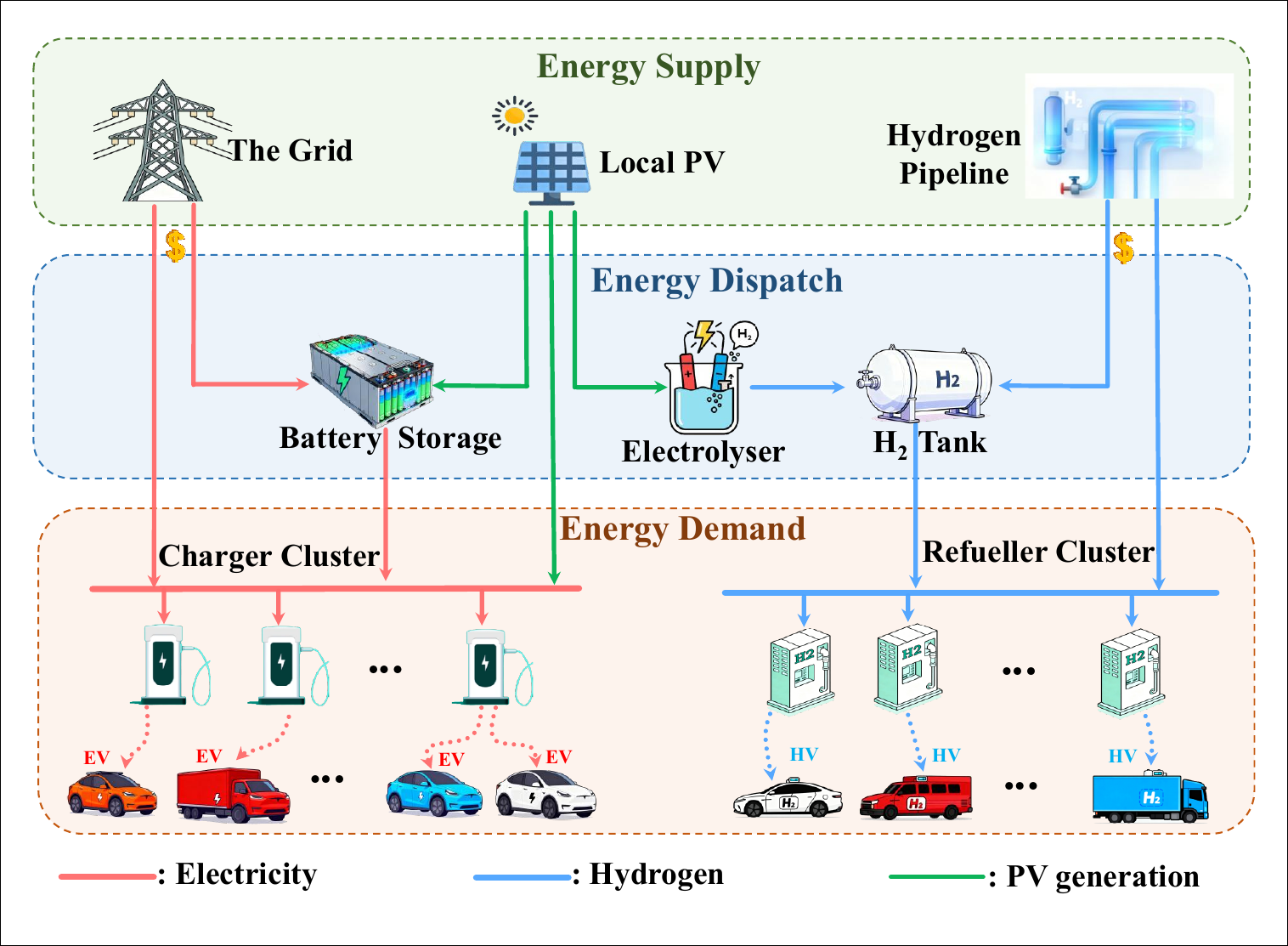}
		%\captionsetup{skip=2.0pt} % 
           \captionsetup{skip=-3 pt}
		\caption{Schematic of the integrated EHTS} 
		\label{fig:1}
	\end{center}
\end{figure*}
 Figure \ref{fig:1} illustrates the proposed integrated EHTS, which couples electricity, hydrogen, and transport networks. The system integrates energy supply (PV generation, grid connection, and hydrogen pipeline), energy storage (battery and hydrogen tank), energy conversion (electrolyser), and vehicle service infrastructures (chargers and refuellers) to support coupled electricity–hydrogen-transport operations. 
 From the transport-service perspective, the EHTS contains multiple chargers and refuellers that dynamically interact with arriving EVs and HVs. Since vehicle arrivals, energy demands, and service durations vary over time, real-time vehicle scheduling is required to efficiently allocate charging and refuelling resources while maintaining service quality. To meet these transport demands in a cost-effective manner, the system integrates multiple interconnected energy components. 
 
On the electricity side, EV charging demand can be supplied by PV generation, battery discharge, and grid electricity. PV generation is always prioritised as the local renewable energy source to satisfy EV charging demand, while any surplus PV generation can either be stored in the battery or converted into hydrogen through the electrolyser. The EHTS is assumed to be connected to the grid through an import-only (unidirectional) connection. Therefore, surplus PV generation cannot be exported to the utility grid. Instead, when the available PV generation exceeds the combined local demand, battery charging, and hydrogen production, the surplus renewable energy is curtailed. Electricity from the grid can also be used to support EV or battery charging when it is economically favourable. The battery thus provides operational flexibility by shifting electricity consumption across different periods, enabling renewable energy utilisation and electricity price arbitrage between low-price and high-price periods. On the hydrogen side, refuelling demand is satisfied through hydrogen sourced from the external hydrogen pipeline and local hydrogen production. To avoid temporal mismatches between the electrolysis process and hydrogen demand, the produced hydrogen is typically not directly supplied to HVs but is first stored in the hydrogen tank for later use when needed. The hydrogen tank thus acts as a buffer between hydrogen supply and refuelling demand, providing temporal flexibility and enabling hydrogen price arbitrage. Through the coordinated operation of vehicle scheduling and energy dispatch, the EHTS enables flexible multi-energy management under heterogeneous transport demands. The operational processes and energy interactions are modelled in the following subsections.

\subsection{PV Modelling}\label{sec:2.1}
A local PV system is integrated into the EHTS to provide renewable electricity. The PV output at time $t$, denoted by $P_t^{\mathrm{pv}} \geq 0$ (kW), is modelled as \cite{li2024cooperative,wu2023integrated}:
\begin{equation}
P_t^{\mathrm{pv}} = 10^{-3}\cdot \eta_t^{\mathrm{pv}} \cdot PR \cdot A \cdot G_t
\label{eq:2.1.1}
\end{equation}
where $A$ (m$^2$) is the effective panel area, $G_t \geq 0$ (W/m$^2$) is the solar irradiance, $PR\in(0,1)$ is the performance ratio accounting for system-level losses, and $\eta_t^{\mathrm{pv}}\in(0,1)$ is the temperature-dependent panel conversion efficiency, expressed as
\begin{equation}
\eta_t^{\mathrm{pv}} = \eta^{\mathrm{ref}} \cdot \left[1+\beta \cdot \left(T_t^{\mathrm{ce}}-T^{\mathrm{ref}}
\right)
\right]
\label{eq:2.1.2}
\end{equation}
where $\eta^{\mathrm{ref}}\in(0,1)$ is the panel efficiency under standard test conditions (1000 W/m$^2$, 298 K), $\beta$ (K$^{-1}$) is the temperature coefficient, $T^{\mathrm{ref}}$ (K) is the reference cell temperature, and $T_t^{\mathrm{ce}}$ (K) is the operating cell temperature, calculated by
\begin{equation}
T_t^{\mathrm{ce}} = T_t^{\mathrm{am}} + G_t \cdot \frac{T^{\mathrm{ce,no}} - T^{\mathrm{am,no}}}{G^{\mathrm{si,no}}}
\label{eq:2.1.3}
\end{equation}
where $T_t^{\mathrm{am}}$ (K) is the ambient temperature, while $T^{\mathrm{am,no}}$ (K), $T^{\mathrm{ce,no}}$ (K), and $G^{\mathrm{si,no}}$ (W/m$^2$) represent the ambient temperature, operating cell temperature, and solar irradiance under nominal operating conditions, respectively.

\subsection{Electrolyser Modelling}  \label{sec:2.2}
A polymer electrolyte membrane (PEM) electrolyser is employed to convert allocated PV power into hydrogen, thereby coupling the electrical and hydrogen networks. The electrolysis process is modelled as \cite{wu2023integrated}:
\begin{equation}
H_t^{el,tk} = \frac{P_t^{pv,el}\cdot\Delta t}{\eta^{el}}
\label{eq:2.2.1}
\end{equation}
where $P_t^{pv,el}$ (kW) is the PV power allocated to the electrolyser, $H_t^{el,tk}\ge0$ (Nm$^3$) is the hydrogen produced during interval $[t,t+\Delta t]$, $\Delta t$ (h) is the interval duration, and $\eta^{el}>0$ (kWh/Nm$^3$) is the electricity consumption required to produce one unit of hydrogen.

\subsection{Battery Modelling} \label{sec:2.3}
The lithium-ion battery is used to store grid electricity and surplus PV generation while supporting EV charging demand. The state-of-charge (SOC) dynamics of the battery are modelled as
\begin{equation}
SOC_{t+1}^{bt} = SOC_t^{bt} -\frac{P_t^{bt,cc} \cdot \Delta t} {E^{bt,cp} \cdot \eta^{bt,dc}} +\frac{P_t^{pv,bt} \cdot \Delta t \cdot \eta^{bt,ch}} {E^{bt,cp}}
+\frac{P_t^{gd,bt} \cdot \Delta t \cdot \eta^{bt,ch}} {E^{bt,cp}}
\label{eq:2.3.1}
\end{equation}
where $SOC_t^{bt} \in [0,1]$ represents the battery SOC, $E^{bt,cp}>0$ (kWh) denotes the battery capacity, $P_t^{bt,cc}\ge0$ (kW) is the electrical power discharged to the charger cluster (EVs), and $P_t^{pv,bt}$ and $P_t^{gd,bt}\ge0$ (kW) are the battery charging power supplied by PV generation and grid electricity, respectively. $\eta^{bt,ch}$ and $\eta^{bt,dc}$ are the battery charging and discharging efficiencies, respectively. The battery charging and discharging powers are limited by the rated power 
\begin{equation}
0 \le P_t^{bt,cc} \le P^{bt,max}
\label{eq:2.3.2}
\end{equation}
\begin{equation}
0 \le P_t^{gd,bt}, \qquad 0 \le P_t^{pv,bt}, \qquad P_t^{gd,bt}+P_t^{pv,bt} \le P^{bt,max}
\label{eq:2.3.3}
\end{equation}
where $P^{bt,max}$ (kW) is the maximum battery charging/discharging power. The  battery's operating mode is subsequently determined by the dispatch action defined in Section \ref{sec:3.2.1}, where positive actions correspond to discharging and negative actions correspond to charging. Furthermore, the battery operation must satisfy SOC limits:
\begin{equation}
P_t^{bt,cc}
\le \frac{\left(SOC_t^{bt} - SOC^{bt,min}\right) \cdot E^{bt,cp} \cdot \eta^{bt,dc}} {\Delta t}
\label{eq:2.3.4}
\end{equation}

\begin{equation}
P_t^{gd,bt} + P_t^{pv,bt} \le \frac{\left( SOC^{bt,max} - SOC_t^{bt} \right) \cdot E^{bt,cp}}
{\eta^{bt,ch} \cdot \Delta t}
\label{eq:2.3.5}
\end{equation}
where $SOC^{bt,min}$ and $SOC^{bt,max}$ denote the minimum and maximum allowable SOC levels, respectively.

\subsection{Hydrogen Tank Modelling} \label{sec:2.4}

A hydrogen tank is adopted to store hydrogen produced by the electrolyser or purchased from the hydrogen pipeline. The SOC dynamics of the hydrogen tank are modelled as
\begin{equation}
SOC_{t+1}^{tk} = SOC_t^{tk} - \frac{ H_t^{tk,rc}}{H^{tk,cp}\cdot\eta^{tk,dc}}
+ \frac{H_t^{el,tk}\cdot\eta^{tk,ch}}{H^{tk,cp}}
+ \frac{H_t^{pl,tk}\cdot\eta^{tk,ch}}{H^{tk,cp}}
\label{eq:2.4.1}
\end{equation}
where $SOC_t^{tk} \in[0,1]$ represents the hydrogen tank SOC,
$H^{tk,cp}>0$ (Nm$^3$) denotes the tank capacity,
$H_t^{tk,rc}\ge0$ (Nm$^3$) is the hydrogen supplied to the refueller cluster (HVs),
and $H_t^{el,tk}\ge0$ and $H_t^{pl,tk}\ge0$ (Nm$^3$) are the hydrogen supplied from the electrolyser and hydrogen pipeline, respectively.
$\eta^{tk,ch}$ and $\eta^{tk,dc}$ are the tank charging and discharging efficiencies, respectively. Similar to the lithium-ion battery model, the operational constraints can be expressed as
\begin{equation}
0 \le H_t^{tk,rc} \le H^{tk,max}
\label{eq:2.4.2}
\end{equation}
\begin{equation}
0 \le H_t^{el,tk}, \qquad 0 \le H_t^{pl,tk},
\qquad H_t^{el,tk}+H_t^{pl,tk} \le H^{tk,max}
\label{eq:2.4.3}
\end{equation}
\begin{equation}
H_t^{tk,rc} \le \left( SOC_t^{tk} - SOC^{tk,min} \right) \cdot H^{tk,cp} \cdot \eta^{tk,dc}
\label{eq:2.4.4}
\end{equation}
\begin{equation}
H_t^{el,tk} + H_t^{pl,tk} \le \frac{\left(SOC^{tk,max}-SOC_t^{tk}\right)\cdot H^{tk,cp}}{\eta^{tk,ch}}
\label{eq:2.4.5}
\end{equation}
where $H^{tk,max}$ (Nm$^3$) denotes the maximum hydrogen amount that can be charged into or discharged from the tank during one dispatch interval \(\Delta t\), and $SOC^{tk,min}$ and $SOC^{tk,max}$ are the minimum and maximum allowable SOC levels of the hydrogen tank, respectively. 

\subsection{Energy Balance Modelling}  \label{sec:2.5}
This subsection formulates the energy balance and supply--demand relationships within the EHTS, involving PV power allocation, EV charging and hydrogen refuelling processes.

\textbf{Energy Balance for PV:} The PV allocation balance among the battery, electrolyser, and charger cluster (EV loads) is expressed as
\begin{equation}
P_t^{pv}\cdot\Delta t = \left(P_t^{pv,bt} + P_t^{pv,el} + P_t^{pv,cc} + P_t^{pv,cur} \right) \cdot \Delta t
\label{eq:2.5.1}
\end{equation}
where $\{P_t^{pv,bt},P_t^{pv,el},P_t^{pv,cc}\}\ge0$ (kW) represent the PV power allocated to the battery, electrolyser, and charger cluster, respectively, while $P_t^{pv,cur}\ge0$ (kW) denotes the curtailed PV power.

\textbf{Energy Balance for Charger Cluster:} Neglecting transmission losses, the supply--demand balance for the charger cluster can be expressed as
\begin{equation}
P_t^{cc,ecs}\cdot\Delta t = \left(P_t^{gd,cc} + P_t^{bt,cc} + P_t^{pv,cc} \right) \cdot \Delta t
\label{eq:2.5.2}
\end{equation}
where $P_t^{cc,ecs}\ge0$ (kW) represents the aggregated EV charging demand from all chargers, and
$P_t^{gd,cc}$ denotes the electricity supplied from the grid. The aggregated charging demand during one simulation interval can be further calculated by

\begin{equation}
P_t^{cc,ecs}\cdot\Delta t = \sum_{p=1}^{N^{ec}} \frac{\tau_{p,t}^{ec}}{\Delta t} \cdot P^{ec,rt} \cdot \Delta t
\label{eq:2.5.3}
\end{equation}
where $N^{ec}$ is the number of chargers, $\tau_{p,t}^{ec}$ (h) represents the occupied time of charger $p$ within interval $[t,t+\Delta t]$, and $P^{ec,rt}$ (kW) denotes the rated charging power of each charger. The ratio $\tau_{p,t}^{ec}/\Delta t\in[0,1]$ therefore represents the charger utilisation during the interval. Here, all chargers are assumed to be identical and operate at their rated charging power whenever occupied.

\textbf{Energy Balance for Refueller Cluster:} The hydrogen balance of the refueller cluster is represented as
\begin{equation}
H_t^{rc,hrs} = H_t^{tk,rc} + H_t^{pl,rc}
\label{eq:2.5.4}
\end{equation}
\begin{equation}
H_t^{rc,hrs} = \sum_{q=1}^{N^{hr}} \frac{\tau_{q,t}^{hr}}{\Delta t} \cdot h^{hr,rt} \cdot \Delta t \label{eq:2.5.5}
\end{equation}
where $H_t^{rc,hrs}\ge0$ (Nm$^3$) represents the aggregated hydrogen refuelling demand of all refuellers, $\tau_{q,t}^{hr}$ (h) denotes the active refuelling duration of refueller $q$ within interval $[t,t+\Delta t]$, and $h^{hr,rt}$ (Nm$^3$/h) is the rated hydrogen refuelling capacity of each refueller. Similar to the charger cluster, all refuellers are assumed to be identical and operate at rated capacities.

\subsection{Vehicle Scheduling Modelling}  \label{sec:2.6}

The scheduling of EVs and HVs within the EHTS is inherently stochastic and dynamic. During each decision interval $t$ with duration $\Delta t$, newly arriving vehicles enter the system, while existing vehicles may continue waiting or receiving service. Each vehicle is characterised by heterogeneous energy demands and service durations, which require real-time scheduling of available charging and refuelling resources.

Let $S_t^{wt,ev}$ and $S_t^{wt,hv}$ denote the waiting sets of EVs and HVs at the beginning of interval $t$, respectively. Each waiting set contains both newly arrived vehicles and those carried over from previous intervals. For each vehicle $i\in S_t^{wt,ev}$ (or $j\in S_t^{wt,hv}$), its characteristics are defined by a charging (or refuelling) demand $E_{i,t}^{ev}$, $E_{j,t}^{hv}>0$ (kWh for EVs and Nm$^3$ for HVs) and a waiting time $w_{i,t}^{ev}$, $w_{j,t}^{hv}\ge0$ (h). Each charger $p\in\mathcal{C}$ and refueller $q\in\mathcal{R}$ operates at a fixed rated service rate of $P^{ec,rt}$ (kW) or $h^{hr,rt}$ (Nm$^3$/h), respectively. The present model assumes homogeneous piles (chargers or refuellers) within each service type: the charger cluster $\mathcal{C}$ contains $N^{ec}$ identical chargers, and the refueller cluster $\mathcal{R}$ contains $N^{hr}$ identical refuellers. The implications of extending the scheduling method to heterogeneous piles are discussed in Section \ref{sec:4.6}.

At every decision instant, the scheduler needs to assign waiting vehicles to available chargers and refuellers. The scheduling process follows two fundamental operational principles: non-preemptive service and within-interval sequential assignment. The former ensures that an ongoing charging or refuelling process cannot be interrupted before completion, while the latter allows a charger or refueller to sequentially serve multiple vehicles within the same interval provided that sufficient service time remains.

\subsubsection{Non-Preemptive Service} \label{sec:2.6.1}

Under the non-preemptive service principle, once a vehicle is connected to a pile, the service proceeds continuously at the pile's rated capacity until the vehicle's demand is satisfied. If the service is not completed within the interval, the vehicle remains connected and continues receiving service in subsequent intervals. 
A pile becomes available for a new vehicle only after the current service has been completed. This rule preserves service continuity and reflects practical charging and refuelling operations. For EV $i$, the charging energy $e_{i,t}^{ev}$ delivered during interval $[t,t+\Delta t]$ depends on its service status and falls into one of the following three mutually exclusive cases: \textit{1)} if EV $i$ is not assigned to any charger during this interval, then $e_{i,t}^{ev}=0$; \textit{2)} if EV $i$ has been connected to a charger before $t$ and continues charging within this interval, then $e_{i,t}^{ev} = \min\!\left(
E_{i,t}^{ev}, P^{ec,rt}\cdot\Delta t\cdot\eta^{ec} \right)$, where $\eta^{ec}\in(0,1)$ represents the charger efficiency; \textit{3)} if EV $i$ starts charging on charger $p$ at an offset time $\tau_{i,p,t}^{ev,s}\in[0,\Delta t)$ (relative to the start of interval $t$), the available charging time is reduced to $\Delta t-\tau_{i,p,t}^{ev,s}$, then $ e_{i,t}^{ev} = \min\!\left( E_{i,t}^{ev}, P^{ec,rt}\cdot(\Delta t-\tau_{i,p,t}^{ev,s})\cdot\eta^{ec} \right)$. Therefore,
\begin{equation}
e_{i,t}^{ev} = \begin{cases} 0, & \text{if not assigned},\\[2mm]
\min\!\left(E_{i,t}^{ev},P^{ec,rt}\cdot\Delta t\cdot\eta^{ec}\right), & \text{if already in service}, \\[2mm]
\min\!\left(E_{i,t}^{ev},P^{ec,rt}\cdot(\Delta t-\tau_{i,p,t}^{ev,s})\cdot\eta^{ec}\right), & \text{if starts charging}.
\end{cases}
\label{eq:2.6.1}
\end{equation}

The remaining charging demand of EV $i$ evolves as
\begin{equation}
E_{i,t+1}^{ev} = E_{i,t}^{ev} - e_{i,t}^{ev}.
\label{eq:2.6.2}
\end{equation}

Accordingly, the accumulated waiting time is updated by
\begin{equation}
w_{i,t+1}^{ev} = \begin{cases}
w_{i,t}^{ev}+\Delta t, & \text{if not assigned}, \\[2mm]
w_{i,t}^{ev}, & \text{if already in service}, \\[2mm]
w_{i,t}^{ev}+\tau_{i,p,t}^{ev,s}, & \text{if starts charging},
\end{cases}
\label{eq:2.6.3}
\end{equation}
where $w_{i,t}^{ev}$ records the cumulative waiting duration experienced before service initiation. 

The refuelling process for HVs is modelled in the same manner. If HV $j$ is assigned to refueller $q$ with a rated service rate $h^{hr,rt}$, the hydrogen delivered during interval $\Delta t$ can be expressed as
\begin{equation}
e_{j,t}^{hv}  = \begin{cases}
0, & \text{if not assigned}, \\[2mm]
\min\!\left(E_{j,t}^{hv},h^{hr,rt}\cdot\Delta t\cdot\eta^{hr}\right), & \text{if already in service}, \\[2mm]
\min\!\left(E_{j,t}^{hv},h^{hr,rt}\cdot(\Delta t-\tau_{j,q,t}^{hv,s})\cdot\eta^{hr}\right), &
\text{if starts refuelling}.
\end{cases}
\label{eq:2.6.4}
\end{equation}

The remaining refuelling demand and accumulated waiting time of HV $j$ evolve as
\begin{equation}
E_{j,t+1}^{hv} = E_{j,t}^{hv} - e_{j,t}^{hv},
\label{eq:2.6.5}
\end{equation}
\begin{equation}
w_{j,t+1}^{hv} = \begin{cases}
w_{j,t}^{hv}+\Delta t, & \text{if not assigned}, \\[2mm]
w_{j,t}^{hv}, & \text{if already in service}, \\[2mm]
w_{j,t}^{hv}+\tau_{j,q,t}^{hv,s}, & \text{if starts refuelling}.
\end{cases}
\label{eq:2.6.6}
\end{equation}

\subsubsection{Within-Interval Sequential Assignment} \label{sec:2.6.2}

Unlike classical one-to-one scheduling that allows each charger or refueller to serve at most one vehicle within a decision interval, the designed scheduler permits sequential assignment of multiple vehicles to the same pile, as long as the sequential service schedule satisfies the non-preemptive constraint and the cumulative occupied time does not exceed the interval duration. This mechanism captures the practical operation in which a pile completes the service of one vehicle and immediately begins serving the next waiting vehicle. For modelling simplicity, the transition time between consecutive vehicles is neglected. The charging duration of EV $i$ on charger $p$ within interval $[t,t+\Delta t]$ depends on its start offset $\tau_{i,p,t}^{ev,s}$ and remaining demand $E_{i,t}^{ev}$, which is expressed as
\begin{equation}
\tau_{i,p,t}^{ev} = \begin{cases}
0, & \text{if not assigned}, \\[2mm]
\min\!\left(\frac{E_{i,t}^{ev}}{\eta^{ec}\cdot P^{ec,rt}},\Delta t \right), & \text{if already in service}, \\[3mm]
\min\!\left( \frac{E_{i,t}^{ev}} {\eta^{ec}\cdot P^{ec,rt}}, \Delta t-\tau_{i,p,t}^{ev,s}
\right), & \text{if starts charging}.
\end{cases}
\label{eq:2.6.7}
\end{equation}

After serving EV $i$, the remaining available time of the charger $p$ within the current interval is
\begin{equation}
\tau_{i,p,t}^{ec,rm} = \Delta t - \tau_{i,p,t}^{ev} - \tau_{i,p,t}^{ev,s}.
\label{eq:2.6.8}
\end{equation}

The remaining available time can subsequently be allocated to the next waiting vehicle. If additional residual time remains after serving the next assigned vehicle, the process continues iteratively until no further vehicles can be accommodated within the interval. Accordingly, the charging energy delivered to the next assigned EV $ii$ is
\begin{equation}
e_{ii,t}^{ev} = \min\!\left(E_{ii,t}^{ev}, P^{ec,rt}\cdot\tau_{i,p,t}^{ec,rm}\cdot\eta^{ec}
\right).
\label{eq:2.6.9}
\end{equation}

The occupancy used in (\ref{eq:2.5.3}) combines the charging durations of ongoing and newly initiated sessions:
\begin{equation}
\tau_{p,t}^{ec}
= \sum_{i\in\mathcal{A}_{p,t}^{ev}} \tau_{i,p,t}^{ev},
\qquad
0 \leq \tau_{p,t}^{ec} \leq \Delta t.
\label{eq:2.6.10}
\end{equation}
Here, $\mathcal{A}_{p,t}^{ev}$ comprises all EVs receiving energy from charger $p$ within $[t,t+\Delta t]$, whether their charging begins during this interval or continues from an earlier one.

Similarly, the service duration of HV $j$ on refueller $q$ is given by
\begin{equation}
\tau_{j,q,t}^{hv} = \begin{cases}
0, & \text{if not assigned}, \\[2mm]
\min\!\left(\frac{E_{j,t}^{hv}}{\eta^{hr}\cdot h^{hr,rt}},\Delta t \right), & \text{if already in service}, \\[3mm]
\min\!\left(\frac{E_{j,t}^{hv}}{\eta^{hr}\cdot h^{hr,rt}},\Delta t-\tau_{j,q,t}^{hv,s}
\right), & \text{if starts refuelling}.
\end{cases}
\label{eq:2.6.11}
\end{equation}

The remaining available time of refueller $q$ after serving HV $j$, the hydrogen delivered to the next assigned HV $jj$, and the cumulative occupied time of refueller $q$ (as in (\ref{eq:2.5.5})) are represented as
\begin{equation}
\tau_{j,q,t}^{hr,rm} = \Delta t - \tau_{j,q,t}^{hv} - \tau_{j,q,t}^{hv,s}
\label{eq:2.6.12}
\end{equation}
\begin{equation}
e_{jj,t}^{hv} = \min\!\left( E_{jj,t}^{hv}, h^{hr,rt}\cdot\tau_{j,q,t}^{hr,rm}\cdot\eta^{hr} \right).
\label{eq:2.6.13}
\end{equation}
\begin{equation}
\tau_{q,t}^{hr} = \sum_{j\in {A}_{q,t}^{hv}} \tau_{j,q,t}^{hv},
\qquad
0 \leq \tau_{q,t}^{hr} \leq \Delta t.
\label{eq:2.6.14}
\end{equation}

The proposed within-interval sequential assignment improves charger and refueller utilisation by allocating remaining service time to subsequent waiting vehicles. Compared with conventional one-vehicle-per-pile scheduling schemes, it reduces resource underutilisation and increases service throughput. The service sequence is constructed at the beginning of each decision interval using the available waiting vehicles and estimated pile completion times. Subsequent vehicles can therefore start service when the preceding service finishes, without waiting for the next decision boundary or requiring a new scheduling decision at each completion event.

\subsection{Hierarchical Objectives}  \label{sec:2.7}

% The operation of the EHTS involves multiple coupled energy flows among the PV system, grid, battery, electrolyser, hydrogen tank, hydrogen pipeline, and charging/refuelling infrastructures. 
Consistent with the hierarchical architecture of the EHTS, the objectives are categorised into a vehicle scheduling objective and an energy dispatch objective. The former aims to improve service quality by reducing vehicle waiting and overflow times, while the latter focuses on promptly satisfying the resulting charging and refuelling demands as well as reducing the operational costs associated with electricity purchasing, hydrogen purchasing, and PV curtailment.

\textbf{Energy Dispatch Objective:} The objective of the energy dispatch is to minimise the energy-supply cost
\begin{equation}
C_t^{ed} = C_t^{gd} + C_t^{pl} + C_t^{pv}
\label{eq:2.7.1}
\end{equation}
where
\begin{equation}
C_t^{gd} = p_t^{elc} \cdot \left( P_t^{gd,cc} + P_t^{gd,bt} \right) \cdot \Delta t \label{eq:2.7.2}
\end{equation}
\begin{equation}
C_t^{pl} = p_t^{h2} \cdot \left( H_t^{pl,rc} + H_t^{pl,tk} \right)
\label{eq:2.7.3}
\end{equation}
\begin{equation}
C_t^{pv} = p_t^{elc} \cdot P_t^{pv,cur} \cdot \Delta t
\label{eq:2.7.4}
\end{equation}
where $C_t^{gd}$ and $C_t^{pl}$ (£) denote the actual expenditure on grid electricity and pipeline hydrogen, respectively, whereas $C_t^{pv}$ (£) represents an  opportunity cost term associated with PV curtailment. The variables $p_t^{elc}$ (£/kWh) and $p_t^{h2}$ (£/Nm³) denote the time-varying electricity and hydrogen prices. In (\ref{eq:2.7.4}), curtailed PV energy is valued at the prevailing electricity purchase price as a proxy for its potential value through local utilisation. This term is a modelling valuation rather than an actual payment or a loss of export revenue, since the grid connection permits imports only. Accordingly, $C_t^{ed}$ combines procurement expenditure and the PV-curtailment opportunity cost term. Unless otherwise stated, “PV curtailment cost” and “PV cost” refer to this opportunity cost term throughout the paper.

\textbf{Vehicle Scheduling Objective:} The objective of the vehicle scheduling is to minimise the total service delay
\begin{equation}
C_t^{vs} = C_t^{wt} + C_t^{ot}
\label{eq:2.7.5}
\end{equation}
where
\begin{equation}
C_t^{wt} = \sum_{i\in S_t^{wt,ev}} \left( w_{i,t+1}^{ev} - w_{i,t}^{ev} \right) + \sum_{j\in S_t^{wt,hv}} \left( w_{j,t+1}^{hv} - w_{j,t}^{hv} \right)
\label{eq:2.7.6}
\end{equation}
\begin{equation}
C_t^{ot} = \sum_{i\in S_t^{wt,ev}} \max \left( 0, w_{i,t+1}^{ev} - w^{ev,max} \right)
+ \sum_{j\in S_t^{wt,hv}} \max \left( 0, w_{j,t+1}^{hv} - w^{hv,max} \right)
\label{eq:2.7.7}
\end{equation}
where $C_t^{wt}$ and $C_t^{ot}$ represent the total waiting-time increment of all vehicles during interval $t$ and the cumulative overflow time beyond the maximum waiting threshold ($w^{ev,max}$ for EVs and $w^{hv,max}$ for HVs), respectively. Here, $C_t^{wt}$ reflects the short-term growth of vehicle waiting time within the current interval, whereas $C_t^{ot}$ captures the accumulated overflow time exceeding the allowable service threshold. Unlike the waiting-time term, the overflow-time term continuously accumulates excessive waiting time, thereby assigning progressively greater importance to persistently delayed vehicles and discouraging long-term service congestion. Since both $C_t^{wt}$ and $C_t^{ot}$ are expressed in the same unit (hours), they are directly aggregated without introducing additional weighting coefficients.

\section{Optimisation Framework}\label{sec:3}
This section presents the proposed hierarchical optimisation framework for the integrated EHTS. 
The framework consists of two sequentially coupled layers: vehicle scheduling determines charging and refuelling assignments, and energy dispatch subsequently optimises supply to satisfy the resulting demands with the objective of minimising operational costs. The framework therefore follows a service-first, demand-driven structure rather than bidirectional coordinated control.

\begin{figure*}[t]
	\begin{center}
        \includegraphics[width=17cm, trim=5 5 5 5, clip]{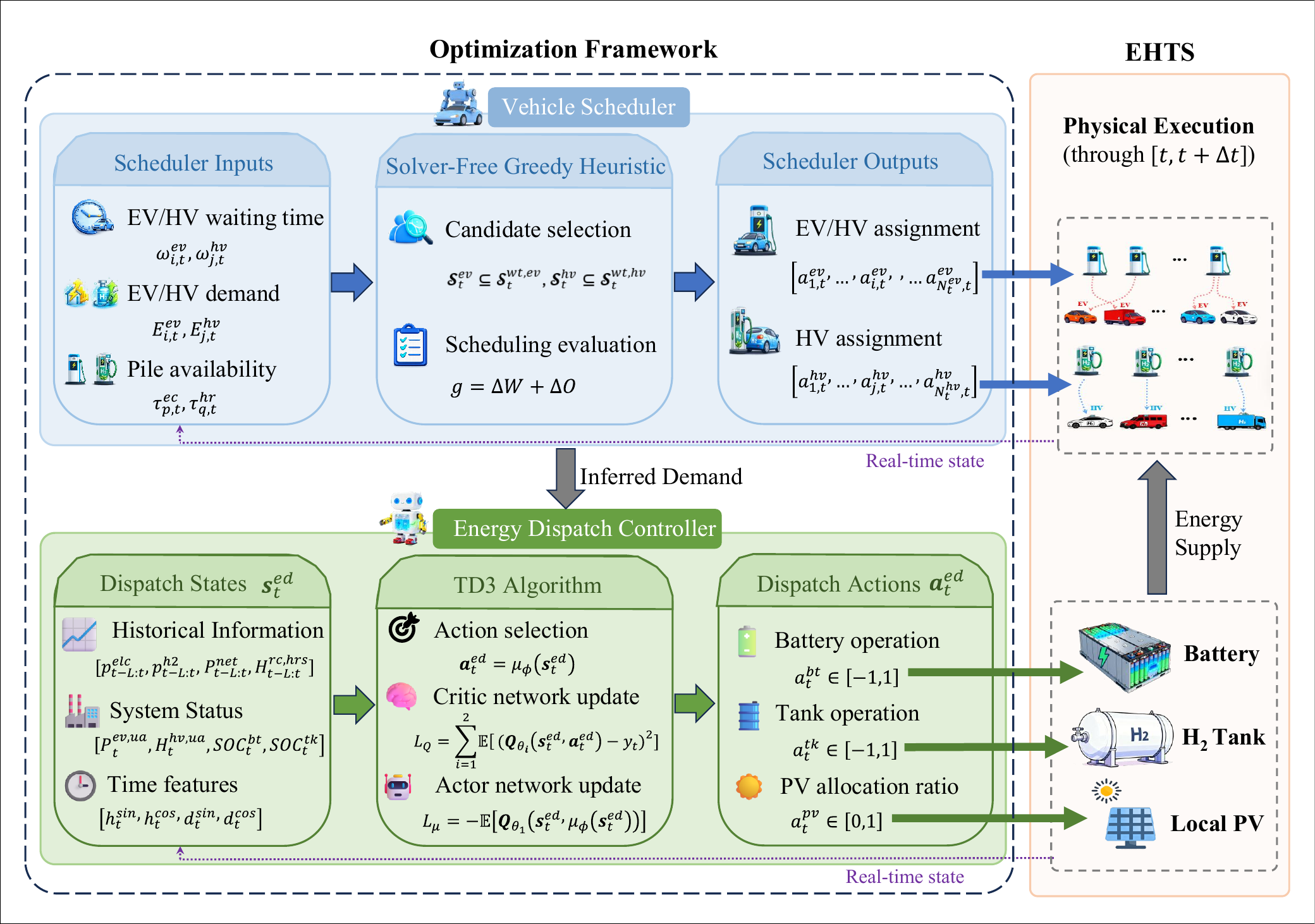}
		%\captionsetup{skip=2.0pt} % 
           \captionsetup{skip=-3 pt}
		\caption{Sequential hierarchical optimisation framework for EHTS} 
		\label{fig:2}
	\end{center}
\end{figure*}

% The scheduling layer determines EV charging and HV refuelling schedules, thereby generating the corresponding electricity and hydrogen demand profiles, whereas the dispatch layer coordinates multiple energy resources to satisfy these demands at minimise operational costs.
Figure \ref{fig:2} illustrates the overall optimisation process. At the beginning of each decision interval, newly arriving EVs and HVs are added to their respective waiting pools. A solver-free greedy heuristic (SFGH) scheduling algorithm then determines feasible charging and refuelling assignments according to vehicle waiting times, charging/refuelling requirements, and pile availability. Based on the resulting assignment decision, the scheduler immediately infers the cluster-level EV charging demand and HV refuelling demand expected to occur during the upcoming interval $[t, t+\Delta t]$. These inferred demands are transferred to the dispatch layer. A TD3-based energy dispatch approach then observes the current system state and generates continuous control actions for battery charging/discharging, hydrogen tank charging/discharging, and surplus PV generation allocation. The generated actions are subsequently projected onto the feasible operating region defined by device ratings, SOC limits, and energy-balance constraints before being executed within the EHTS. 
Through this sequential, demand-driven interaction, the vehicle-scheduling layer determines the charging and refuelling demand to be handled by the downstream energy-dispatch layer, while the dispatch layer reduces operational costs conditioned on the resulting energy demands.

\subsection{Vehicle Scheduling Layer}  \label{sec:3.1}
To achieve real-time vehicle scheduling under dynamic arrival conditions, the SFGH algorithm operates through three stages within each decision interval: candidate selection, feasible-assignment evaluation and generation, and demand inference. Based on the current EV/HV waiting pools, vehicle energy demands, and charger/refueller availability, a limited set of high-priority vehicles is first selected as scheduling candidates. Feasible vehicle--pile assignments are then evaluated using a greedy scoring criterion, and the best assignment is iteratively accepted until no additional feasible assignment can be accommodated within the current interval. Finally, the obtained charging and refuelling schedules are converted into cluster-level transport demand signal and transferred to the energy-dispatch layer. Unlike optimisation-based approaches that repeatedly solve mathematical programs at every decision interval, the proposed SFGH directly constructs scheduling decisions through greedy evaluation, resulting in substantially lower computational complexity and making it suitable for real-time operation. The scheduling objective follows the service-delay minimisation criterion defined in (\ref{eq:2.7.5}), which jointly minimises the waiting-time increment and accumulated overflow time.

To avoid evaluating all waiting vehicles at every scheduling interval, only a limited number of high-priority vehicles are retained as scheduling candidates. Specifically, at most $N^{ev}$ EVs and $N^{hv}$ HVs are selected from the waiting sets during each scheduling interval. Vehicles are ranked lexicographically according to larger overflow time, longer waiting time, and smaller charging/refuelling demand, in that order. The resulting candidate subsets are denoted as $S_t^{ev}\subseteq S_t^{wt,ev}$ and $S_t^{hv}\subseteq S_t^{wt,hv}$. The SFGH determines charging and refuelling assignments using
\begin{equation}
a_t^{vs}= \left[a_{1,t}^{ev},\ldots,a_{i,t}^{ev},\ldots,a_{N_t^{ev},t}^{ev};a_{1,t}^{hv},\ldots,a_{j,t}^{hv},\ldots,a_{N_t^{hv},t}^{hv}
\right],
\label{eq:3.1.1}
\end{equation}
\begin{equation}
a_{i,t}^{ev} \in \{0,1,\ldots,p,\ldots,N^{ec}\}, 
\qquad
a_{j,t}^{hv} \in \{0,1,\ldots,q,\ldots,N^{hr}\}.
\label{eq:3.1.2}
\end{equation}
where $a_{i,t}^{ev}=p$ and $a_{j,t}^{hv}=q$ indicate that EV $i$ is assigned to charger $p$ and HV $j$ is assigned to refueller $q$, respectively. A value of zero indicates that the corresponding vehicle remains in the waiting queue during interval $t$. 

Assignment feasibility is determined according to the remaining available service time of chargers and refuellers, as defined in (\ref{eq:2.6.8}) and (\ref{eq:2.6.12}). Among all feasible assignments, the SFGH evaluates the immediate scheduling benefit using
\begin{equation}
g = \Delta W + \Delta O,
\label{eq:3.1.3}
\end{equation}
where $\Delta W$ and $\Delta O$ denote the reductions in waiting time and overflow time achieved by the assignment, respectively. The feasible assignment with the highest score is selected first. Once an assignment is accepted, the corresponding charger or refueller state is updated according to the service completion and residual-time update rules described in Section \ref{sec:2.6}. The SFGH then re-evaluates the remaining feasible assignments based on the updated resource availability. This process is repeated until no further feasible assignment can be accommodated within the current decision interval.

Once the scheduling assignments have been determined, the occupancies of chargers and refuellers during the upcoming interval $[t,t+\Delta t]$ are known before physical execution. Consequently, the cluster-level EV charging demand and HV refuelling demand expected over this interval can be inferred immediately without waiting for the interval to finish. The inferred EV charging demand and HV refuelling demand are calculated according to (\ref{eq:2.5.3}) and (\ref{eq:2.5.5}), respectively, and transferred to the dispatch layer for energy-management decisions. 

\subsection{Energy Dispatch Layer}  \label{sec:3.2}

Given the charging and refuelling demands inferred by the vehicle scheduling layer, the energy dispatch layer coordinates the battery storage, hydrogen tank, and surplus PV generation during the same interval. Its objective is to reduce the overall operational costs of the EHTS while satisfying the real-time charging and refuelling demands. 
% Unlike the scheduling layer, which operates on a variable-size vehicle set, the dispatch layer deals with aggregated energy flows and therefore can be formulated as a continuous-control decision problem with a fixed-dimensional state space.
At each decision epoch, the dispatch agent observes the current energy-system condition and determines continuous control actions. In this research, a Twin Delayed Deep Deterministic Policy Gradient (TD3) agent is employed to learn an effective dispatch policy.

\subsubsection{Dispatch Problem Representation} \label{sec:3.2.1}

To enable effective learning, the dispatch problem is represented through carefully designed state, action, and reward definitions that capture both the system's physical characteristics and economic objectives.

\textbf{State Space:} 
The performance of DRL-based energy management strongly depends on the quality of the state representation. Many existing studies incorporate forecasted information into the state space to provide future awareness. However, recent research \cite{yao2025unified} has shown that forecast-based state representations are beneficial only when the uncertainty associated with exogenous data remains sufficiently low. As the level of uncertainty increases, forecast errors can propagate through the learning process, ultimately degrading the performance of the learned control policy. Under such conditions, history record-based state representations become a more robust alternative, as they avoid the adverse effects of inaccurate forecasts while still capturing the temporal dynamics of the system.

In integrated EHTSs, multiple stochastic processes coexist, including PV generation, electricity prices, hydrogen prices, and aggregated charging/refuelling demand. In particular, the aggregated EV charging and HV refuelling demand jointly depend on stochastic vehicle arrivals and heterogeneous service requirements, resulting in compounded uncertainty that is difficult to forecast accurately. Given these characteristics, this research work adopts a history record-based state representation rather than relying on forecasted information. The dispatch state is defined as
\begin{equation}
s_t^{ed} = \left[p_{t-L:t}^{elc},p_{t-L:t}^{h2},P_{t-L:t}^{um},H_{t-L:t}^{rc,hrs},P_t^{ev,ua},H_t^{hv,ua},SOC_t^{bt},SOC_t^{tk},h_t^{sin},h_t^{cos},d_t^{sin},d_t^{cos}\right]
\label{eq:3.2.1}
\end{equation}
where $p_{t-L:t}^{elc}$, $p_{t-L:t}^{h2}$, $P_{t-L:t}^{um}$, and $H_{t-L:t}^{rc,hrs}$ denote the historical sequences of electricity price, hydrogen price, unmet power (defined as $P_t^{um} = P_t^{cc,ecs} - P_t^{pv}$), and aggregated hydrogen-refuelling demand over the look-back horizon $[t-L,t]$, respectively. Furthermore, $P_t^{ev,ua}$ (kW) and $H_t^{hv,ua}$ (Nm$^3$) denote the inferred aggregated EV charging demand and HV refuelling demand, respectively, remaining in the waiting pool after the vehicle-scheduling process at current interval. The variables $h_t^{sin}$, $h_t^{cos}$, $d_t^{sin}$, and $d_t^{cos}$ are sinusoidal time encodings representing the hour-of-day and day-of-week information, which preserve the cyclical nature of time and facilitate temporal pattern learning. By incorporating both current operating conditions and historical system evolution, the dispatch state captures the temporal dependencies required for coordinated energy management under uncertain transport demands.

\textbf{Action Space:} The dispatch agent determines the operation of the battery storage, hydrogen tank, and PV generation allocation through a three-dimensional continuous action vector
\begin{equation}
a_t^{ed} = \left[a_t^{bt},a_t^{tk},a_t^{pv}\right]
\label{eq:3.2.2}
\end{equation}
\begin{equation}
a_t^{bt}\in[-1,1],
\qquad
a_t^{tk}\in[-1,1],
\qquad
a_t^{pv}\in[0,1].
\label{eq:3.2.3}
\end{equation}

The battery action $a_t^{bt}$ specifies the battery charging/discharging request and is converted into a power request according to
\begin{equation}
P_t^{bt,req} = a_t^{bt}\cdot P^{bt,max}.
\label{eq:3.2.4}
\end{equation}

Here, a positive value of $P_t^{bt,req}$ (kW) indicates battery discharge to supply EV charging demand (i.e., $P_t^{bt,cc} = P_t^{bt,req}$), whereas a negative value indicates battery charging (i.e., $(P_t^{gd,bt} + P_t^{pv,bt}) = - P_t^{bt,req}$). Therefore, the sign of $a_t^{bt}$ uniquely determines the battery operating mode, inherently preventing simultaneous charging and discharging. 

Similarly, the hydrogen-tank action $a_t^{tk}$ determines the tank charging/discharging request and is converted into a hydrogen-flow request as
\begin{equation}
H_t^{tk,req} = a_t^{tk}\cdot H^{tk,max}
\label{eq:3.2.5}
\end{equation}
where positive $H_t^{tk,req}$ corresponds to hydrogen release from the tank to support HV refuelling demand (i.e., $H_t^{tk,rc} = H_t^{tk,req}$), whereas negative value indicates hydrogen-tank charging (i.e., $(H_t^{el,tk} + H_t^{pl,tk})= -H_t^{tk,req}$). 

Surplus PV generation is allocated to battery charging and hydrogen production (via the electrolyser) according to
\begin{equation}
P_t^{pv,bt} = a_t^{pv} \cdot  max (P_t^{pv}-P_t^{cc,ecs},0),
\label{eq:3.2.6}
\end{equation}
\begin{equation}
P_t^{pv,el} =\left(1-a_t^{pv}\right)\cdot max (P_t^{pv}-P_t^{cc,ecs},0).
\label{eq:3.2.7}
\end{equation}

PV generation always supplies EV charging demand first, with surplus PV (if available) allocated according to $a_t^{pv}$ in (\ref{eq:3.2.6}) and (\ref{eq:3.2.7}), subject to constraints (\ref{eq:2.3.3}), (\ref{eq:2.3.5}), (\ref{eq:2.4.3}), and (\ref{eq:2.4.5}). PV can be allocated to battery charging or electrolysis only when the corresponding storage device is in charging mode. When neither the battery nor the hydrogen tank is in charging mode, all surplus PV is curtailed, as the unidirectional grid connection permits electricity imports only. The battery and hydrogen-tank requests in (\ref{eq:3.2.4}) and (\ref{eq:3.2.5}) are scaled according to the feasible charging/discharging margins defined by (\ref{eq:2.3.2})–(\ref{eq:2.3.5}) and (\ref{eq:2.4.2})–(\ref{eq:2.4.5}), respectively. For devices in charging mode, the allocated surplus PV (if available) is used first. Battery discharge is limited to the EV demand remaining after PV supply, while hydrogen-tank discharge is limited to the HV demand. Grid electricity and pipeline hydrogen then cover the residual vehicle demand and unmet charging requests, satisfying the balances in (\ref{eq:2.5.1}), (\ref{eq:2.5.2}), and (\ref{eq:2.5.4}). Finally, the realised flows then update the storage SOCs through (\ref{eq:2.3.1}) and (\ref{eq:2.4.1}).

\textbf{Reward Function:} The EHTS involves multiple cost components with significantly different numerical scales. In particular, hydrogen refuelling demands, hydrogen prices, and hydrogen-storage capacities are generally much larger than their electricity counterparts. Consequently, directly using the operational cost objective in (\ref{eq:2.7.1}) as the reward would cause hydrogen-related terms to dominate the learning signal, making it difficult for the agent to learn effective battery scheduling and PV generation allocation behaviours. To address this issue, this work adopts the direct-response reward formulation proposed in \cite{yao2025unified}, which has proven effective for DRL problems with heterogeneous reward scales. Rather than evaluating the operational costs, the reward measures the immediate economic benefit achieved by the dispatch action relative to a predefined baseline operation. The dispatch reward is defined as\begin{equation}
r_t^{ed} = B_t^{gd} + B_t^{pl} + B_t^{pv},
\label{eq:3.2.8}
\end{equation}
\begin{equation}
B_t^{gd} = C_t^{gd,base} - C_t^{gd},
\label{eq:3.2.9}
\end{equation}
\begin{equation}
B_t^{pl} = C_t^{pl,base} - C_t^{pl},
\label{eq:3.2.10}
\end{equation}
\begin{equation}
B_t^{pv} = C_t^{pv,base} - C_t^{pv}.
\label{eq:3.2.11}
\end{equation}

Here, $B_t^{gd}$, $B_t^{pl}$, and $B_t^{pv}$ represent the economic benefits achieved through battery, hydrogen tank, and PV generation controls, respectively. The baseline costs $C_t^{gd,base}$, $C_t^{pl,base}$, and $C_t^{pv,base}$ correspond to a no-control operation in which all unmet electricity demand ($P_t^{um}$) is supplied by the grid, all hydrogen refuelling demand ($H_t^{rc,hrs}$) is supplied by the hydrogen pipeline, and any surplus PV generation is curtailed.

By evaluating cost reduction relative to the corresponding baseline operation rather than the absolute operational cost, the proposed reward removes the basic uncontrollable cost components shared by both strategies.
% As a result, the resulting benefit terms become more balanced, alleviating scale imbalances among objectives and preventing hydrogen-related costs from dominating the learning process.
This provides a more informative learning signal and facilitates policy learning, while remaining fully aligned with the operational objective defined in Section \ref{sec:2.7}.

\subsubsection{TD3-Based Dispatch} \label{sec:3.2.2}

TD3 adopts an actor--critic architecture consisting of one deterministic actor network and two independent critic networks \cite{fujimoto2018addressing}. The actor generates a continuous dispatch action according to the observed system state 
\begin{equation} 
a_t^{ed} = \mu_{\phi}\left(s_t^{ed}\right), \label{eq:3.2.12} 
\end{equation} 
where $\mu_{\phi}(\cdot)$ denotes the  deterministic actor policy parameterised by $\phi$. The two critic networks estimate the action-value function 
\begin{equation} 
Q_{\theta_1}\left(s_t^{ed},a_t^{ed}\right), \qquad Q_{\theta_2}\left(s_t^{ed},a_t^{ed}\right), 
\label{eq:3.2.13} 
\end{equation} 
where $\theta_1$ and $\theta_2$ denote the parameters of the two critic networks. To improve robustness against function-approximation errors, TD3 adopts the target-policy smoothing strategy during training by adding a small clipped Gaussian noise to the target action generated by the target actor network, 
\begin{equation} 
a_{t+1}^{ed} = \mu_{\phi'}\left(s_{t+1}^{ed}\right) +\varepsilon, \qquad \varepsilon\sim \mathrm{clip} \left( \mathcal{N}(0,\sigma_a^2), -c_a, c_a \right), 
\label{eq:3.2.15} 
\end{equation} 
where $\phi'$ denotes the parameters of the target actor network, $\varepsilon$ is the smoothing noise sampled from a clipped Gaussian distribution, $\sigma_a$ is the standard deviation of the Gaussian noise, and $c_a$ is the clipping threshold. To alleviate the overestimation bias commonly observed in deterministic policy-gradient methods, TD3 also employs clipped double-Q learning. The target value is computed as 
\begin{equation} 
y_t = r_t^{ed} + \gamma \min_{i=1,2} Q_{\theta_i'} \left( s_{t+1}^{ed}, a_{t+1}^{ed} \right), 
\label{eq:3.2.14} 
\end{equation} 
where $\gamma$ is the discount factor, and $\theta_i'$ denotes the parameters of the $i$-th target critic network. The critic networks are updated by minimising the temporal-difference loss 
\begin{equation} 
L_Q = \sum_{i=1}^{2} \mathbb{E} \left[ \left( Q_{\theta_i} \left( s_t^{ed}, a_t^{ed} \right) - y_t \right)^2 \right]. 
\label{eq:3.2.16} 
\end{equation}

To further improve learning stability, TD3 updates the actor less frequently than the critic networks. The actor objective is 
\begin{equation} 
L_{\mu} = - \mathbb{E} \left[ Q_{\theta_1} \left( s_t^{ed}, \mu_{\phi} \left( s_t^{ed} \right) \right) \right]. 
\label{eq:3.2.17} 
\end{equation}

After each delayed actor update, the target actor and target critic networks are softly updated according to 
\begin{equation} 
\phi' \leftarrow \rho\phi + (1-\rho)\phi', \qquad
\theta_i' \leftarrow \rho\theta_i + (1-\rho)\theta_i', \qquad i=1,2,
\label{eq:3.2.18} 
\end{equation} 
where $\rho\in(0,1)$ is the soft-update coefficient. Through repeated interactions with the EHTS environment, the TD3 agent gradually learns a dispatch policy that maximises the cumulative reward. 
% Since the reward formulation is directly derived from the operational objective in (\ref{eq:2.7.1}), the learned policy simultaneously improves renewable-energy utilisation, battery and hydrogen-storage operation, and overall economic performance of the integrated electric--hydrogen-transport system.

\section{Simulation and Analysis}\label{sec:4}

This section presents a comprehensive evaluation of the proposed hierarchical optimisation framework. It first introduces the simulation setup and system configuration, followed by comprehensive performance, comparative, and ablation analyses. Generalisation under variations in transport demand and PV penetration is then evaluated, followed by a discussion of practical deployment considerations.

\subsection{Simulation Setup and System Configuration} \label{sec:4.1}

This work adopts a hybrid dataset comprising real-world meteorological and electricity data together with representative synthetic transport and hydrogen data. Further details regarding the dataset construction are provided in \ref{app:1}. Four year-long datasets were generated, each containing 35040 time steps at a 15-min (i.e., $\Delta t=0.25$h) temporal resolution. One dataset was used for training, whereas the remaining three datasets were held out as independent test sets. According to \cite{wu2023integrated}, the PV system was configured as $\eta^{ref}=0.22$, $\beta=-0.001\%$/K, $T^{ref}=298$ K, $T^{ce,no}=318$ K, $T^{am,no}=293$ K, $PR=0.85$, $G^{si,no}=800$ W/m$^2$, and $A=1000$ m$^2$. 
Based on \cite{wu2023integrated,knosala2021hybrid}, the electrolyser, battery storage, and hydrogen tank were parameterised as $\eta^{el}=5.0$ kWh/Nm$^3$, $\eta^{bt,ch}=\eta^{bt,dc}=0.95$, $SOC^{bt,min}=0.10$, $SOC^{bt,max}=0.95$, $\eta^{tk,ch}=\eta^{tk,dc}=0.995$, $SOC^{tk,min}=0.10$, and $SOC^{tk,max}=0.95$.
The battery storage capacity was set to $E^{bt,cp}=900$ kWh, with a rated charging/discharging power limit of $P^{bt,max}=450$ kW. The hydrogen tank capacity was set to $H^{tk,cp}=2400$ Nm$^3$, with a rated hydrogen charging/discharging flow limit of 1200 Nm$^3$/h. For the 15-min dispatch interval, this corresponds to a maximum hydrogen transfer amount of \(H^{tk,max}=300\) Nm$^3$ per interval. For the vehicle service system, the charger and refueller configurations were selected with reference to realistically deployed infrastructures and existing studies \cite{sawant2024dc,atabay2024design,zhao2025system}. The charger cluster $\mathcal{C}$ consists of $N^{ec}=5$ EV chargers with $P^{ec,rt}=90$ kW, while the refueller cluster $\mathcal{R}$ consists of $N^{hr}=3$ hydrogen refuellers with $h^{hr,rt}=400$ Nm$^3$/h. The charger and refueller delivery efficiencies were both set to $\eta^{ec}=\eta^{hr}=0.99$, while the maximum waiting-time thresholds were set to $w^{ev,max} = 2\Delta t$ for EVs and $w^{hv,max} = 3\Delta t$ for HVs.

For the vehicle scheduler design, at each decision interval, at most $N^{ev}=10$ EVs and $N^{hv}=6$ HVs were selected to construct the candidate subsets $S_t^{ev}$ and $S_t^{hv}$, respectively. For the TD3 controller, the historical input length was set to (L=6) with a stride of 4 for each exogenous and load signal, corresponding to the previous 6 hours of information. Both the actor and critic consisted of two fully connected hidden layers with 256 neurons and a learning rate of ($3\times10^{-4}$). The discount factor and target-network update coefficient were set to $\gamma=0.995$ and $\rho=0.005$, respectively. The TD3 was trained for 8,000 rollout epochs, each containing 672 consecutive steps (one week). A replay buffer of 500,000 transitions and a mini-batch size of 256 were employed. The target-policy smoothing parameters were configured as $\sigma_a=0.15$ and $c_a=0.3$. During training, Gaussian exploration noise with the same standard deviation ($0.15$) was added to the actor output, while uniformly random actions were employed during the first 2,000 interaction steps. The state variables were normalised to the range of $[0,1]$ using min--max normalisation, and the rewards were scaled by a factor of 0.1 to improve training stability. The initial SOCs of the battery and hydrogen tank were independently sampled during training stage from a uniform distribution over $[0.10,0.95]$ at each epoch reset. During evaluation, both storage devices were initialised at an SOC of $0.10$. No fixed terminal SOC target or cyclic SOC constraint was imposed. For cost accounting, the remaining battery energy and hydrogen inventory at the end of the evaluation horizon were valued at the terminal electricity and hydrogen prices, respectively. These residual values were deducted from the corresponding electricity and hydrogen costs, and all reported total costs include this adjustment. All simulations were implemented in Python using PyTorch and run on a Windows 11 workstation with an Intel i7-14700F CPU, 32~GB RAM, and an NVIDIA RTX 5070 Ti GPU.

\subsection{Performance Analysis}\label{sec:4.2}

\begin{figure}[]
	\begin{center}
        \includegraphics[width=17cm, trim=5 5 5 5, clip]{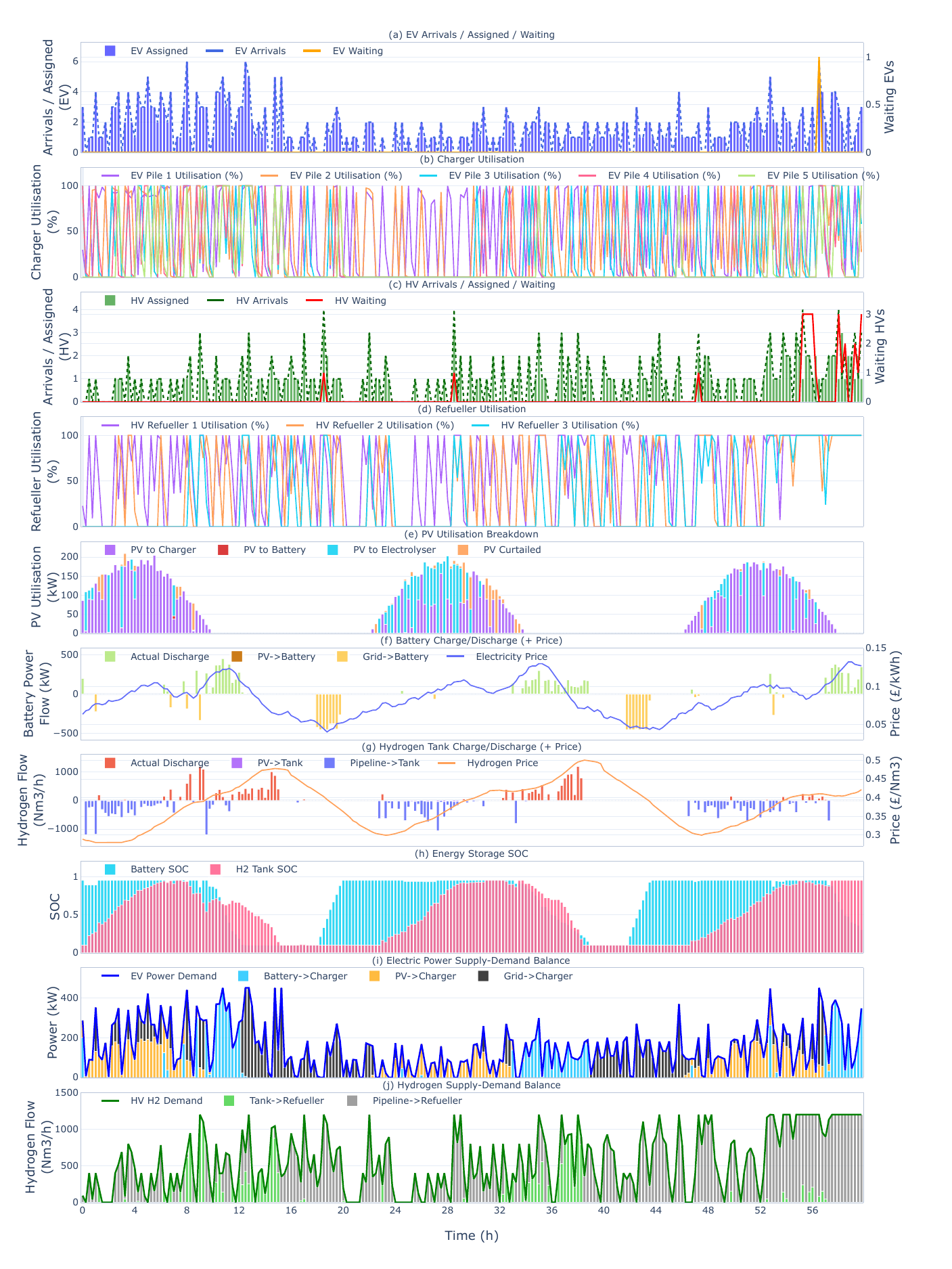}
		%\captionsetup{skip=2.0pt} % 
           \captionsetup{skip=-3 pt}
		\caption{Short-term simulation results for EHTS} 
		\label{fig:3}
	\end{center}
\end{figure}

Fig. \ref{fig:3} illustrates the integrated operation of the EHTS under the proposed hierarchical optimisation framework over a representative 60-hour period.
A dynamic dashboard is available in the GitHub repository \cite{ehts_dashboard}, providing further insights into the long-term behaviour of the system. The results demonstrate the coordinated behaviour of the vehicle-scheduling layer and energy-dispatch layer under uncertain EV/HV demands, renewable-energy generation, and time-varying energy prices. 

The first observation is that the SFGH scheduling algorithm effectively maintains service quality while achieving high infrastructure utilisation. As shown in Figs. \ref{fig:3}(a) and \ref{fig:3}(c), for both EVs and HVs, the numbers of assigned vehicles closely follow the arrival processes, indicating that the SFGH is capable of accommodating dynamic transport demand in real time. Although temporary congestion occasionally occurs during peak-demand periods, the accumulated queues are rapidly dissipated rather than persisting over extended durations. This behaviour indicates that the scheduling layer successfully prevents long-term service bottlenecks despite stochastic arrivals and heterogeneous service requirements. Furthermore, the charger and refueller utilisation profiles shown in Figs. \ref{fig:3}(b) and \ref{fig:3}(d) indicate that all piles are actively utilised and frequently operate at full capacity. The workload is distributed across different chargers and refuellers without persistent overloading or idleness, demonstrating efficient resource allocation and balanced infrastructure utilisation.

The second observation is that the TD3-based optimisation approach successfully allocates renewable generation while managing battery-storage and hydrogen-tank operations, as shown in Figs. \ref{fig:3}(e)–\ref{fig:3}(h). As defined in Section \ref{sec:3.2.1}, PV generation is prioritised to satisfy EV charging demand, resulting in the PV-to-charger pathway accounting for the largest share of PV utilisation, as shown in Fig. \ref{fig:3}(e). When surplus PV generation is available, the dispatch policy tends to allocate it to hydrogen production through the electrolyser rather than battery charging. Such a strategy is economically reasonable because the hydrogen prices are generally higher than  electricity prices, allowing surplus renewable energy to generate greater economic value through hydrogen production. Also, PV generation curtailment generally occurs when the battery storage and hydrogen tank approach their maximum SOC limits or when storage discharge is economically preferred during high-price periods. The battery-storage and hydrogen-tank operations shown in Figs. \ref{fig:3}(f) and \ref{fig:3}(g) further reveal clear price-responsive behaviours. Both storage systems charge during low-price periods and discharge during high-price periods, demonstrating successful energy arbitrage learned by the TD3 agent. It is also worth noting that although surplus PV generation is preferentially converted into hydrogen, the amount of hydrogen produced by the electrolyser, in Fig. \ref{fig:3}(g), remains relatively small compared with pipeline purchases. This is partly because PV generation is prioritised to satisfy EV charging demand, thereby limiting the PV energy available for hydrogen production.
% This reflects the economic trade-off learned by the TD3 agent between renewable-energy utilisation and storage arbitrage. 
Furthermore, the energy storage SOC trajectories in Fig \ref{fig:3}(h) confirm that both battery and tank devices actively participate in system balancing while remaining within their allowable operating ranges. 

The final observation is that the coordinated operation of renewable generation and energy storage substantially reduces dependence on external energy supplies. As shown in Figs. \ref{fig:3}(i) and \ref{fig:3}(j), EV charging demand is jointly supplied by PV generation, battery discharge, and grid electricity import, while HV refuelling demand is satisfied by both pipeline hydrogen and hydrogen-tank discharge. 
% Although pipeline hydrogen remains the dominant supply source, the battery and hydrogen tank effectively buffer demand fluctuations and partially substitute external energy purchases. 
Compared with conventional charging and refuelling stations that rely entirely on grid electricity and pipeline hydrogen, the integration of PV generation, battery storage, electrolytic hydrogen production, and hydrogen storage substantially increases renewable-energy utilisation and reduces operational costs. 

Overall, the results demonstrate that the proposed hierarchical framework effectively optimises vehicle scheduling and multi-energy dispatch, thereby achieving low service delays, high charging-resource utilisation, enhanced renewable-energy utilisation, and reduced operational costs. To further quantitatively evaluate the proposed framework, comparative studies and ablation analyses are presented for both the vehicle-scheduling and energy-dispatch layers in the following subsections.

\subsection{Comparative Studies for Vehicle Scheduling Layer}\label{sec:4.3}
For the vehicle-scheduling layer, the proposed SFGH algorithm was compared with four rule-based scheduling strategies. All four rule-based algorithms follow the same two-stage scheduling structure while preserving the non-preemptive service and within-interval sequential assignment introduced in Section \ref{sec:2.6}. First, vehicles in the EV and HV waiting pools are ranked according to a prescribed priority rule. Second, each selected vehicle is assigned to the charger or refueller with the earliest predicted finish time, provided that the assignment is feasible within the current decision interval. Consequently, the compared rule-based algorithms differ only in the vehicle-selection rule, whereas the pile-assignment procedure remains identical. Specifically, FIFO with Earliest-Finish assignment (FIFO+EF) follows the original arrival order of vehicles and therefore represents a conventional first-in-first-out (FIFO) scheduling strategy.  Demand-Arrival with Earliest-Finish assignment (DA+EF) prioritises vehicles with longer accumulated waiting times and uses smaller service demand as a tie-breaker. Least-slack with earliest-finish assignment (LS+EF) prioritises vehicles with the smallest remaining slack, where the slack is defined as the waiting-time threshold minus the current waiting time. Consequently, vehicles closer to violating the waiting-time threshold are served first. Shortest-processing-time with earliest-finish assignment (SPT+EF) prioritises vehicles with shorter estimated charging or refuelling durations, with accumulated waiting time used as a secondary criterion, thereby favouring throughput maximisation. In addition, an ablation variant denoted as SFGH--WISA was considered to evaluate the contribution of the designed within-interval sequential assignment (WISA). SFGH--WISA employs the same vehicle-selection rule as SFGH but replaces WISA with conventional one-to-one scheduling.

\begin{figure*}[t]
    \centering
    \setlength{\fboxrule}{0.1pt}
    \setlength{\fboxsep}{0pt}
    \begin{subfigure}[b]{0.325\textwidth}
        % \centering
        \fcolorbox{gray}{white}{\includegraphics[width=\textwidth]{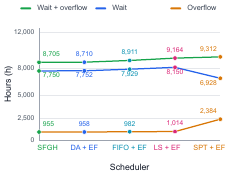}}
        \caption{Wait, overflow, and total hours}
        \label{fig:4}
    \end{subfigure}
    \hfill
    \begin{subfigure}[b]{0.325\textwidth}
        % \centering
        \fcolorbox{gray}{white}{\includegraphics[width=\textwidth]{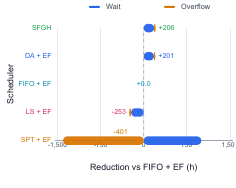}}
        \caption{Wait--overflow reduction decomposition}
        \label{fig:5}
    \end{subfigure}
    \hfill
    \begin{subfigure}[b]{0.325\textwidth}
        % \centering
        \fcolorbox{gray}{white}{\includegraphics[width=\textwidth]{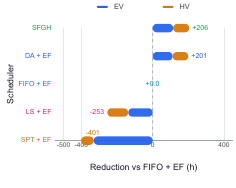}}
        \caption{EV--HV reduction decomposition}
        \label{fig:6}
    \end{subfigure}
    \caption{Test-average performance comparison of different vehicle-scheduling methods, with reductions measured relative to FIFO+EF}
    \label{fig:456}
\end{figure*}

The resulting performance comparison across the training dataset and three test datasets is reported in Table \ref{tab:1}, where the value represents the sum of cumulative waiting and overflow hours for all vehicles over the one-year evaluation horizon. The most notable observation is the performance of SFGH--WISA. Compared with all other scheduling strategies, SFGH--WISA produces substantially larger service delays across both the training and testing datasets, with total delays more than ten times higher than those of the other methods. At the adopted 15-min decision interval, restricting each pile to a single assignment can leave part of the interval unused after the current service completion, even when other vehicles are waiting. WISA uses this remaining time to serve subsequent vehicles, improving pile utilisation and limiting the accumulation of queues. The substantial delay reduction therefore demonstrates the benefit of within-interval pile reuse at the temporal resolution considered; the magnitude of this improvement may change with the decision-interval length. Given this result, the subsequent analysis focuses on the other five scheduling strategies that incorporate WISA. SFGH consistently achieves the lowest service delay across the training dataset and all test datasets, indicating that the proposed SFGH generalises well across different stochastic arrival and demand realisations. On the test average, SFGH reduces the total delay to 8,705 h, outperforming FIFO+EF, DA+EF, LS+EF, and SPT+EF, whose corresponding annual accumulated delays are 8,911 h, 8,710 h,  9,164 h, and 9,312 h, respectively. A more intuitive comparison of the test-average scheduling performance is provided in Fig. \ref{fig:456}(a). Although DA+EF is the closest competing strategy, SFGH still achieves the lowest total value across every dataset. 
\begin{table*}[t]
\centering
\caption{Results of different vehicle scheduling algorithms}
\label{tab:1}
\setlength{\tabcolsep}{10pt}
\renewcommand{\arraystretch}{0.9}
\resizebox{0.8\textwidth}{!}{%
\begin{tabular}{lllllllll}
\toprule
\multicolumn{1}{c}{\multirow{2}{*}{Dataset}} &
\multicolumn{1}{c}{\multirow{2}{*}{Algorithm}} &
\multicolumn{3}{c}{All vehicles (h)} &
\multicolumn{2}{c}{EV (h)} &
\multicolumn{2}{c}{HV (h)} \\
\cmidrule(lr){3-5}
\cmidrule(lr){6-7}
\cmidrule(lr){8-9}
& &
Total & Wait & Overflow &
Wait & Overflow &
Wait & Overflow \\
\midrule
Train & SFGH      & 8677   & 7786  & 891   & 3490  & 311   & 4296  & 580  \\
      & FIFO + EF & 8880   & 7962  & 918   & 3595  & 322   & 4367  & 596  \\
      & DA + EF   & 8685   & 7790  & 895   & 3493  & 314   & 4297  & 581  \\
      & LS + EF   & 9132   & 8179  & 953   & 3721  & 332   & 4458  & 621  \\
      & SPT + EF  & 9334   & 6981  & 2353  & 3229  & 1051  & 3752  & 1302 \\
      & SFGH-WISA & 105359 & 58669 & 46690 & 52625 & 44052 & 6044  & 2638 \\[4pt]
Test 1 & SFGH      & 8791   & 7762  & 1029  & 3821  & 557   & 3941  & 472  \\
       & FIFO + EF & 9006   & 7950  & 1056  & 3929  & 568   & 4021  & 488  \\
       & DA + EF   & 8797   & 7765  & 1032  & 3823  & 560   & 3942  & 472  \\
       & LS + EF   & 9256   & 8167  & 1089  & 4051  & 580   & 4116  & 509  \\
       & SPT + EF  & 9363   & 6943  & 2420  & 3495  & 1289  & 3448  & 1131 \\
       & SFGH-WISA & 107587 & 59669 & 47918 & 53934 & 45439 & 5735  & 2479 \\[4pt]
Test 2 & SFGH      & 8883   & 7938  & 945   & 3457  & 281   & 4481  & 664  \\
       & FIFO + EF & 9083   & 8112  & 971   & 3557  & 291   & 4555  & 680  \\
       & DA + EF   & 8887   & 7940  & 947   & 3458  & 282   & 4482  & 665  \\
       & LS + EF   & 9354   & 8347  & 1007  & 3694  & 303   & 4653  & 704  \\
       & SPT + EF  & 9520   & 7076  & 2444  & 3182  & 1038  & 3894  & 1406 \\
       & SFGH-WISA & 105351 & 58489 & 46862 & 51766 & 43641 & 6723  & 3221 \\[4pt]
Test 3 & SFGH      & 8441   & 7549  & 892   & 3534  & 357   & 4015  & 535  \\
       & FIFO + EF & 8643   & 7725  & 918   & 3637  & 368   & 4088  & 550  \\
       & DA + EF   & 8446   & 7551  & 895   & 3535  & 359   & 4016  & 536  \\
       & LS + EF   & 8882   & 7936  & 946   & 3757  & 377   & 4179  & 569  \\
       & SPT + EF  & 9052   & 6764  & 2288  & 3253  & 1088  & 3511  & 1200 \\
       & SFGH-WISA & 112503 & 62131 & 50372 & 56118 & 47607 & 6013  & 2765 \\[4pt]
Test Avg. & SFGH      & 8705  & 7750  & 955   & 3604  & 398   & 4146  & 557  \\
          & FIFO + EF & 8911  & 7929  & 982   & 3708  & 409   & 4221  & 573  \\
          & DA + EF   & 8710  & 7752  & 958   & 3605  & 400   & 4147  & 558  \\
          & LS + EF   & 9164  & 8150  & 1014  & 3834  & 420   & 4316  & 594  \\
          & SPT + EF  & 9312  & 6928  & 2384  & 3310  & 1138  & 3618  & 1246 \\
          & SFGH-WISA & 108480 & 60096 & 48384 & 53939 & 45562 & 6157 & 2822 \\
\bottomrule
\end{tabular}%
}
\end{table*}

Figs. \ref{fig:456}(b) and \ref{fig:456}(c) further decompose the test-average performance differences relative to FIFO+EF, which represents a commonly adopted queue-based scheduling strategy. As shown in Fig. \ref{fig:456}(b), SFGH achieves simultaneous improvements in both waiting and overflow components. DA+EF exhibits a similar but slightly weaker improvement, whereas LS+EF and SPT+EF degrade overall performance compared to FIFO + EF. The majority of the gain achieved by SFGH originates from waiting-time reduction, while overflow is also effectively controlled. This behaviour is expected because waiting hours constitute the dominant portion of the total service delay, whereas overflow hours accumulate only after the waiting-time threshold is exceeded. In contrast, SPT+EF reduces waiting time more aggressively but causes a substantially larger increase in overflow, resulting in a deterioration of overall performance. Fig. \ref{fig:456}(c) further shows that the improvement achieved by SFGH is distributed across both EVs and HVs rather than being concentrated in a single vehicle class. These results demonstrate that SFGH achieves the best overall performance while maintaining a balanced distribution of improvements.

\subsection{Comparative Studies for Energy Dispatch Layer}\label{sec:4.4}

\begin{figure*}[!b]
    \centering
    \setlength{\fboxrule}{0.1pt}
    \setlength{\fboxsep}{0pt}

    \begin{subfigure}[t]{0.49\textwidth}
        \centering
        \fcolorbox{gray}{white}{\includegraphics[width=\textwidth]{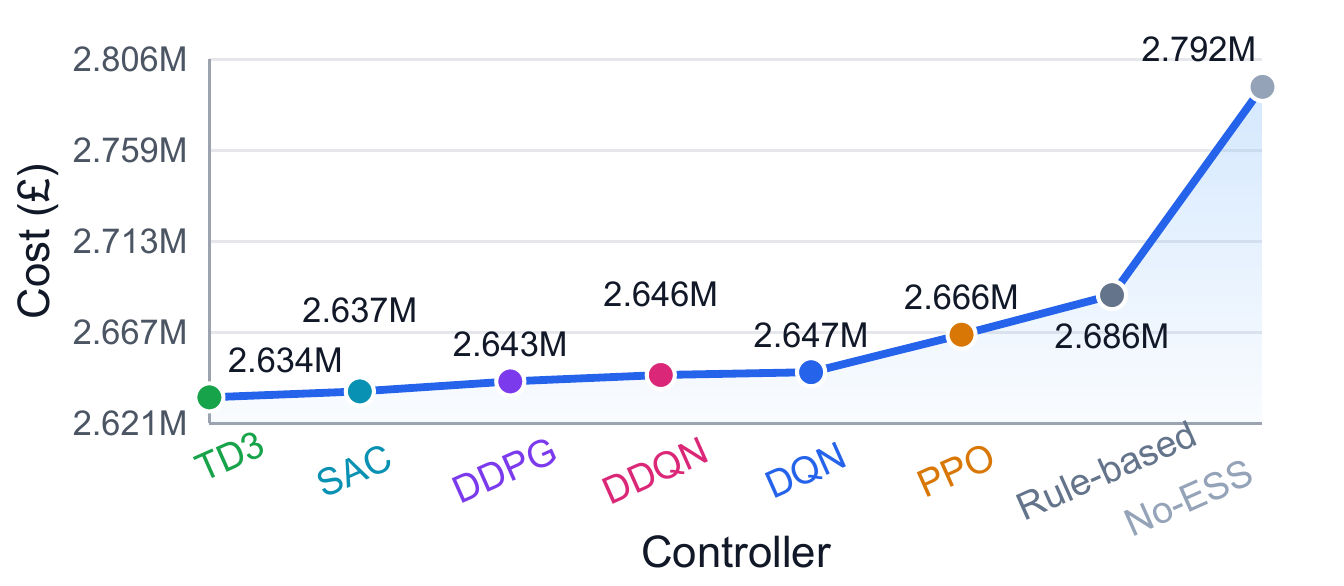}}
        \caption{Total annual operational cost}
        \label{fig:7}
    \end{subfigure}
    \hfill
    \begin{subfigure}[t]{0.49\textwidth}
        \centering
        \fcolorbox{gray}{white}{\includegraphics[width=\textwidth]{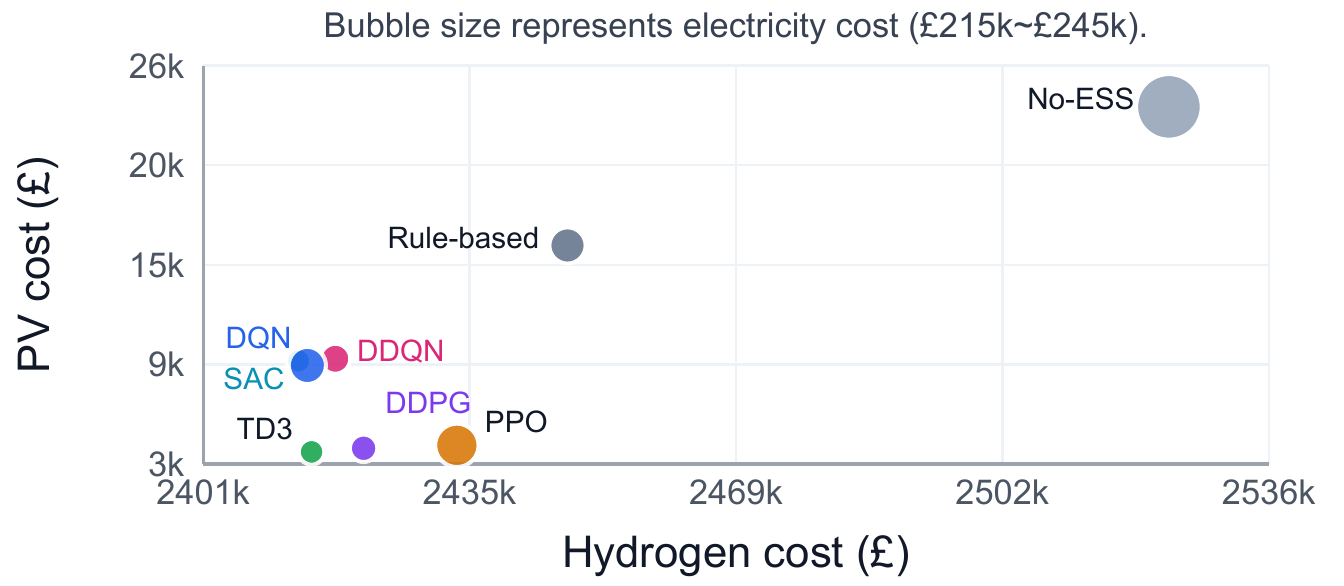}}
        \caption{Total cost trade-off}
        \label{fig:8}
    \end{subfigure}

    \vspace{0.8em}

    \begin{subfigure}[t]{0.49\textwidth}
        \centering
        \fcolorbox{gray}{white}{\includegraphics[width=\textwidth]{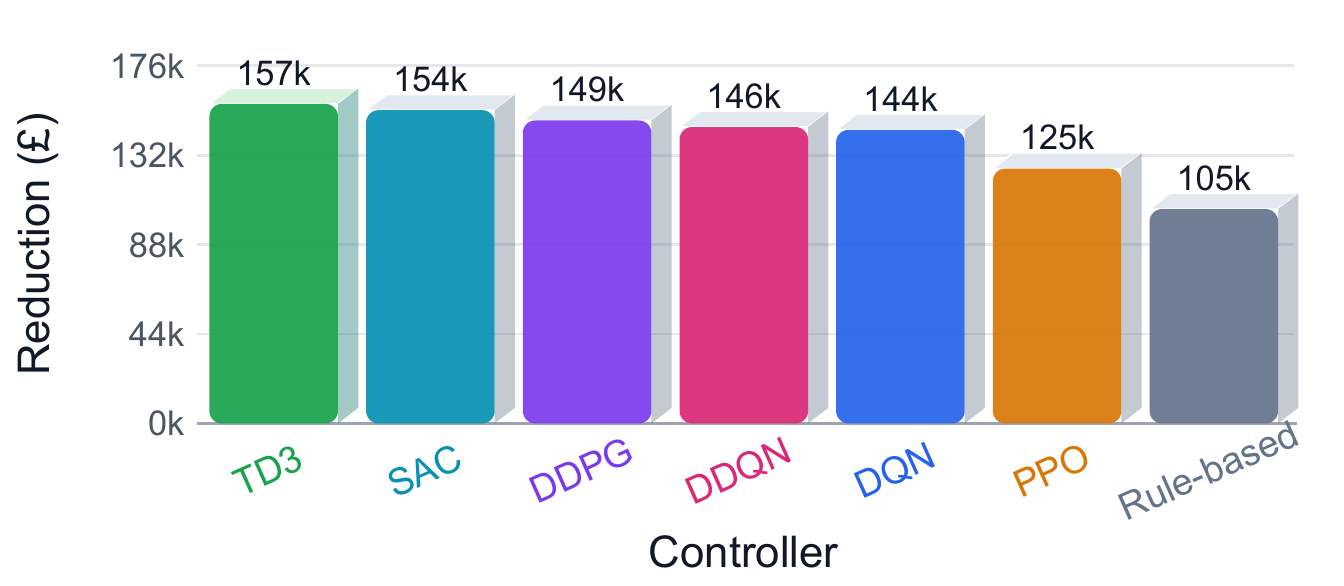}}
        \caption{Cost reduction compared to the No-ESS baseline}
        \label{fig:9}
    \end{subfigure}
    \hfill
    \begin{subfigure}[t]{0.49\textwidth}
        \centering
        \fcolorbox{gray}{white}{\includegraphics[width=\textwidth]{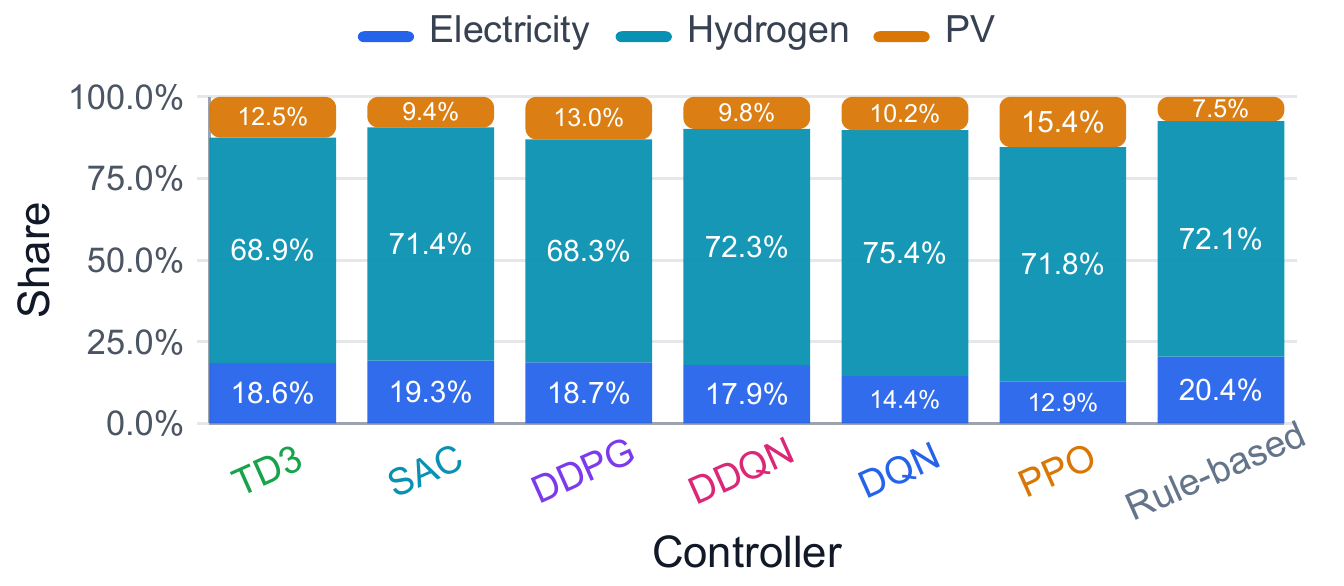}}
        \caption{Cost-reduction contribution share}
        \label{fig:10}
    \end{subfigure}

    \caption{Test-average performance comparison of different ESS control methods, with reductions measured relative to No-ESS}
    \label{fig:7-10}
\end{figure*}

For the energy-dispatch layer, the proposed TD3-based optimisation was compared with five DRL approaches and two non-learning strategies. The DRL approaches include DQN, DDQN, DDPG, PPO, and SAC, covering both discrete- and continuous-action reinforcement-learning paradigms. To ensure a fair comparison, all DRL methods employed the same state representation, reward formulation, and simulation environment defined in Section \ref{sec:3.2.1}. Following the implementation in \cite{yao2025unified}, DQN and DDQN operate on a discretised dispatch action space, where battery-storage and hydrogen-tank actions are represented by five operating levels (full discharge, half discharge, idle, half charge, and full charge), while PV generation allocation is represented by three discrete choices corresponding to battery-priority allocation, hydrogen-priority allocation, and direct curtailment. In contrast, DDPG, PPO, and SAC directly learn continuous dispatch actions in the same manner as TD3. In addition, two non-learning strategies were considered. The first is a rule-based ESS dispatch strategy. PV generation is also first used to meet EV charging demand. Any surplus PV generation is then allocated to battery charging whenever the battery is available to charge; otherwise, it is diverted to the electrolyser for hydrogen production. Any surplus PV generation that cannot be accommodated by either the battery or the electrolyser is curtailed. Battery-storage and hydrogen-tank charging/discharging decisions are determined using rolling electricity- and hydrogen-price thresholds, where the lower one-third and upper two-thirds quantiles are adopted as the charging and discharging thresholds, respectively. The second is a No-ESS strategy, where battery and hydrogen-tank dispatch are disabled, thereby representing an integrated station without active energy-storage utilisation. All energy-dispatch methods were evaluated under the proposed SFGH scheduler, as it achieved the best scheduling performance in the preceding comparison. The resulting annual costs are summarised in Table \ref{tab:2}, and the test-average comparison of total costs is illustrated in Fig. \ref{fig:7-10}(a). 

\begin{table*}[t]
\centering
\caption{Annual costs of different approaches for energy dispatch layer}
\label{tab:2}
\setlength{\tabcolsep}{5.5pt}
\renewcommand{\arraystretch}{0.9}
\resizebox{0.9\textwidth}{!}{%
\begin{tabular}{llllllllll}
\toprule
\multicolumn{1}{c}{\multirow{2}{*}{Dataset}} &
\multicolumn{1}{c}{\multirow{2}{*}{Cost item (\pounds)}} &
\multicolumn{8}{c}{Approach} \\
\cmidrule(lr){3-10}
& & TD3 & SAC & DDPG & PPO & DQN & DDQN & Rule-based & No-ESS \\
\midrule
\multirow{4}{*}{Train}
& Total       & 2635755 & 2639549 & 2643063 & 2668407 & 2647655 & 2647027 & 2687326 & 2793928 \\
& Electricity & 215155 & 214673 & 216312 & 227872 & 223115 & 217724 & 222438 & 244441 \\
& Hydrogen    & 2416733 & 2415328 & 2422315 & 2436459 & 2415736 & 2419282 & 2449097 & 2525807 \\
& PV          & 3867 & 9548 & 4436 & 4076 & 8804 & 10021 & 15791 & 23680 \\[4pt]

\multirow{4}{*}{Test 1}
& Total       & 2624829 & 2627826 & 2633310 & 2656730 & 2638213 & 2636203 & 2676261 & 2782377 \\
& Electricity & 217595 & 217112 & 218869 & 230625 & 225927 & 220148 & 225142 & 246698 \\
& Hydrogen    & 2403239 & 2401482 & 2410250 & 2421748 & 2403228 & 2406328 & 2435602 & 2511972 \\
& PV          & 3995 & 9232 & 4191 & 4357 & 9058 & 9727 & 15517 & 23707 \\[4pt]

\multirow{4}{*}{Test 2}
& Total       & 2664508 & 2668515 & 2672764 & 2694833 & 2677028 & 2675616 & 2716641 & 2822471 \\
& Electricity & 213913 & 213473 & 215515 & 226556 & 222283 & 217284 & 221420 & 243162 \\
& Hydrogen    & 2446448 & 2445665 & 2453270 & 2464304 & 2445808 & 2449479 & 2479551 & 2555676 \\
& PV          & 4147 & 9377 & 3979 & 3973 & 8937 & 8853 & 15670 & 23633 \\[4pt]

\multirow{4}{*}{Test 3}
& Total       & 2613882 & 2616103 & 2621547 & 2647010 & 2626174 & 2625535 & 2665511 & 2769853 \\
& Electricity & 215269 & 214827 & 216507 & 228737 & 223644 & 218640 & 223205 & 244543 \\
& Hydrogen    & 2395154 & 2392768 & 2401014 & 2413876 & 2394188 & 2397996 & 2426669 & 2502185 \\
& PV          & 3459 & 8508 & 4026 & 4397 & 8342 & 8899 & 15637 & 23125 \\[4pt]

\multirow{4}{*}{Test Avg.}
& Total       & 2634406 & 2637481 & 2642540 & 2666190 & 2647138 & 2645785 & 2686138 & 2791567 \\
& Electricity & 215592 & 215137 & 216964 & 228639 & 223951 & 218691 & 223256 & 244801 \\
& Hydrogen    & 2414947 & 2413305 & 2421511 & 2433309 & 2414408 & 2417934 & 2447274 & 2523278 \\
& PV          & 3867 & 9039 & 4065 & 4242 & 8779 & 9160 & 15608 & 23488 \\
\bottomrule
 \multicolumn{10}{l}{\footnotesize Note: Electricity and hydrogen costs are procurement expenditures net of terminal inventory credits; PV cost represents the }\\
  \multicolumn{10}{l}{\footnotesize curtailment opportunity cost, and total cost is the sum of these three components. Also apply to subsequent tables.}\\
\end{tabular}%
}
\end{table*}

A clear performance hierarchy can be observed among the compared approaches. The No-ESS case yields the highest operational cost, followed by the rule-based strategy, whereas all DRL-based approaches achieve substantially lower costs, demonstrating the effectiveness of DRL for integrated energy optimisation. Within the DRL family, the continuous-control methods (TD3, SAC, and DDPG) achieve the lower operational costs, outperforming the discretised-action approaches (DQN and DDQN). Such behaviour indicates that directly optimising continuous dispatch actions is advantageous for the energy-dispatch problem considered in this study. Although PPO also operates in a continuous action space, its performance remains less competitive. This may be because it adopts an on-policy learning paradigm with lower sample efficiency, which limits learning effectiveness in the long-horizon EHTS environment. Among all approaches, TD3 achieves the lowest operational cost, reducing the test-average annual accumulated cost to £2.634 million.
% Its superior performance can be attributed to the twin-critic architecture and delayed policy-update mechanism, both of which improve training stability and mitigate value overestimation during continuous-action optimisation.

Figs. \ref{fig:7-10}(b)--\ref{fig:7-10}(d) provide further insight into the economic performance of these approaches. As shown in Table \ref{tab:2} and Fig. \ref{fig:7-10}(b), TD3 achieves the lowest test-average total cost among the evaluated approaches. Although SAC achieves lower electricity and hydrogen costs, TD3’s lower PV-curtailment cost more than offsets these differences, resulting in a lower total cost. In contrast, the No-ESS strategy performs worst in all three dimensions, as shown in the upper-right region with the largest bubble. The remaining DRL approaches exhibit different optimisation preferences. For example, DQN and SAC achieve relatively lower hydrogen costs, whereas DDPG and PPO are more effective at reducing PV generation curtailment costs. Moreover, hydrogen expenditure clearly dominates the overall operational cost, accounting for approximately £2.401--2.536 million annually, compared with £0.215--0.245 million for electricity purchasing and only £0.003--0.026 million for PV generation curtailment. Consequently, improvements in hydrogen management have the greatest impact on overall economic performance, which is economically intuitive given the substantially higher price of hydrogen relative to electricity on an energy-equivalent basis. Fig. \ref{fig:7-10}(c) further quantifies the annual cost reduction relative to the No-ESS case. TD3 delivers the largest cost reduction of approximately £157k per year, followed closely by SAC (£154k), whereas the rule-based approach achieves only £105k. This indicates that the economic value of energy storage systems cannot be fully realised through fixed operational heuristics and instead requires adaptive learning-based dispatch strategies. 

The contribution analysis in Fig. \ref{fig:7-10}(d) shows that hydrogen-related savings account for the largest share of the achieved cost reduction across all approaches, contributing approximately 68\%--75\% of the total benefit. Electricity-cost reduction contributes around 12\%--20\%, while PV-curtailment cost reduction accounts for the remaining 7.5\%--15\%. This distribution is consistent with the cost structure shown in Fig. \ref{fig:7-10}(b), where hydrogen expenditure dominates the overall operational cost. Although both electricity-cost and PV-generation-curtailment reductions are valued using the same electricity price (Section \ref{sec:2.7}), the latter is constrained by the availability of surplus PV generation, whereas battery dispatch can exploit temporal energy shifting more flexibly. Consequently, electricity-cost reduction contributes a larger share of the overall benefit than PV-generation-curtailment reduction.

Having demonstrated the advantage of continuous over discrete action spaces in Table \ref{tab:2}, further ablation experiments were conducted based on TD3 to evaluate the effects of the state representation and reward formulation, as summarised in Table \ref{tab:3}. TD3$_{r,cost}$ replaces the proposed benefit-based reward in (\ref{eq:3.2.8}) with the raw cost formulation in (\ref{eq:2.7.1}). TD3$_{s,0}$ uses only the current electricity price, hydrogen price, unmet power demand, and aggregated hydrogen-refuelling demand as state variables, without any historical information. TD3$_{s,l}$ ($l=3,9,12$) denotes historical window lengths of $L=l$ in the state representation. The TD3 corresponds to the employed configuration above with $L=6$. As shown in the table, replacing the benefit-based reward with the raw cost formulation leads to the largest performance degradation, increasing the test-average annual operational cost to £2.75 million. Removing historical information also degrades performance, whereas all history-enhanced variants achieve lower costs. Among these configurations, TD3$_{s,6}$ delivers the best overall performance and is therefore adopted in the final controller design.

\begin{table*}[t]
\centering
\caption{Annual costs of different ablation approaches for energy dispatch layer}
\label{tab:3}
\setlength{\tabcolsep}{7pt}
\renewcommand{\arraystretch}{0.9}
\resizebox{0.8\textwidth}{!}{%
\begin{tabular}{llllllll}
\toprule
\multicolumn{1}{c}{\multirow{2}{*}{Dataset}} &
\multicolumn{1}{c}{\multirow{2}{*}{Cost item (\pounds)}} &
\multicolumn{6}{c}{Approach} \\
\cmidrule(lr){3-8}
& & TD3 & TD3$_{r,cost}$ & TD3$_{s,0}$ & TD3$_{s,3}$ & TD3$_{s,9}$ & TD3$_{s,12}$ \\
\midrule

\multirow{4}{*}{Train}
& Total       & 2635755 & 2752641 & 2645046 & 2639011 & 2639588 & 2637514 \\
& Electricity & 215155 & 239452 & 222120 & 215352 & 214606 & 215162 \\
& Hydrogen    & 2416733 & 2495844 & 2418211 & 2419362 & 2419296 & 2418599 \\
& PV          & 3867 & 17345 & 4715 & 4297 & 5686 & 3753 \\[4pt]

\multirow{4}{*}{Test 1}
& Total       & 2624829 & 2742293 & 2639033 & 2628372 & 2630334 & 2628150 \\
& Electricity & 217595 & 242132 & 223795 & 218098 & 216834 & 217758 \\
& Hydrogen    & 2403239 & 2482636 & 2409782 & 2405840 & 2407234 & 2406316 \\
& PV          & 3995 & 17525 & 5456 & 4434 & 6266 & 4076 \\[4pt]

\multirow{4}{*}{Test 2}
& Total       & 2664508 & 2779624 & 2679327 & 2667652 & 2670280 & 2667648 \\
& Electricity & 213913 & 238614 & 219792 & 214292 & 213664 & 213821 \\
& Hydrogen    & 2446448 & 2523775 & 2453980 & 2448928 & 2450555 & 2449770 \\
& PV          & 4147 & 17235 & 5555 & 4432 & 6061 & 4057 \\[4pt]

\multirow{4}{*}{Test 3}
& Total       & 2613882 & 2730810 & 2627622 & 2617003 & 2619195 & 2616015 \\
& Electricity & 215269 & 240567 & 220998 & 215821 & 214811 & 215206 \\
& Hydrogen    & 2395154 & 2473326 & 2401272 & 2397478 & 2398770 & 2397308 \\
& PV          & 3459 & 16917 & 5352 & 3704 & 5614 & 3501 \\[4pt]

\multirow{4}{*}{Test Avg.}
& Total       & 2634406 & 2750910 & 2648660 & 2637675 & 2639936 & 2637271 \\
& Electricity & 215592 & 240438 & 221528 & 216070 & 215103 & 215595 \\
& Hydrogen    & 2414947 & 2493246 & 2421678 & 2417415 & 2418853 & 2417798 \\
& PV          & 3867 & 17226 & 5454 & 4190 & 5980 & 3878 \\
\bottomrule
% \multicolumn{8}{l}{\footnotesize Note: Minor discrepancies ($\pm$1) may occur due to rounding.}\\
\end{tabular}%
}
\end{table*}

\subsection{Framework-Level Comparison and Computational Performance}
\label{sec:framework_comparison}

The energy-dispatch comparisons in Table~\ref{tab:2} all use SFGH for vehicle scheduling. To extend the heuristic scheduling comparisons, SFGH+TD3 is further compared with PPO+TD3 and TD3+TD3, which use learning-based optimisation methods (PPO and TD3) for vehicle scheduling while retaining TD3 for energy dispatch. Given the relatively small cost difference between TD3 and SAC in Table~\ref{tab:2}, SFGH+SAC is also re-evaluated here through multiple independent training runs.

The PPO- and TD3-based schedulers optimise vehicle-service priorities through reward-driven policy learning. Their actor networks output continuous priority scores for candidate vehicles, and sorting these scores determines the vehicle service order. Following this learned order, each vehicle is assigned to the earliest available compatible pile while preserving non-preemptive service and within-interval sequential assignment. Both schedulers use the vehicle and pile information described in Section~\ref{sec:3.1} as their states and the negative service-delay objective in (\ref{eq:2.7.5}), $r_t^{vs}=-C_t^{vs}$, as their reward. All actor and critic networks contain two fully connected hidden layers of 256 neurons each. The TD3 scheduler adopts the same network architecture and hyperparameter settings as TD3-based energy dispatch in Section~\ref{sec:4.1}. For PPO, the actor and critic learning rates are $5\times10^{-5}$ and $10^{-4}$, respectively, with the remaining hyperparameters following \cite{schulman2017proximal}.

\begin{figure*}[htbp]
\centering

\begin{subfigure}[t]{0.49\textwidth}
    \centering
    \includegraphics[width=\linewidth]{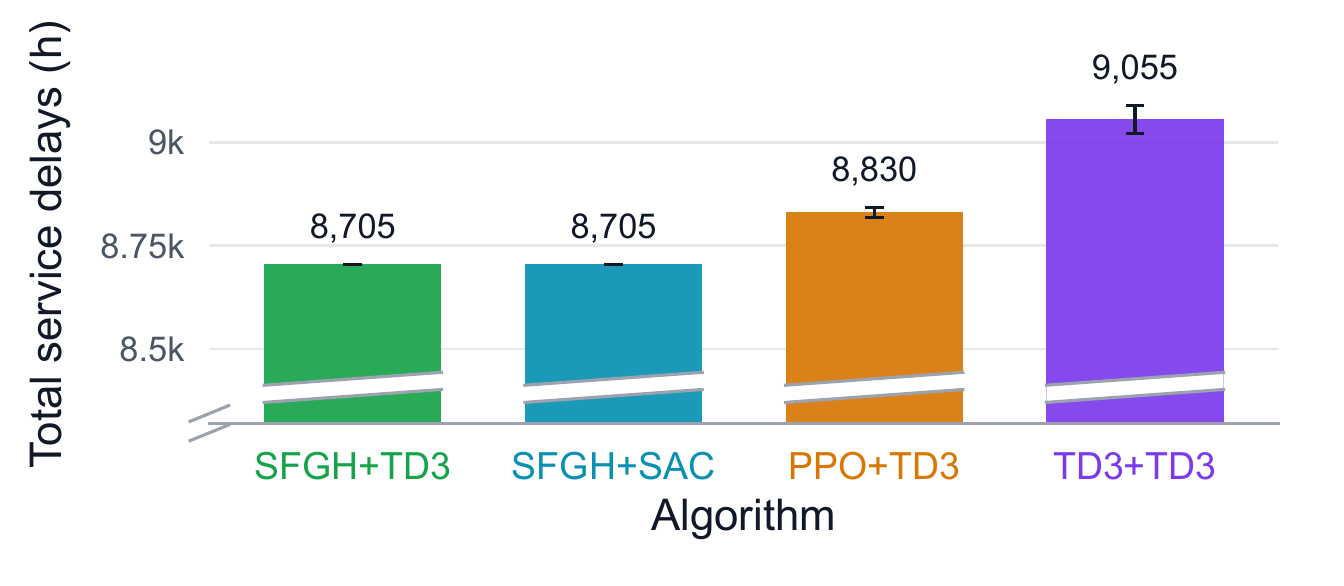}
    \caption{Total service delay.}
    \label{fig:framework_delay}
\end{subfigure}
\hfill
\begin{subfigure}[t]{0.49\textwidth}
    \centering
    \includegraphics[width=\linewidth]{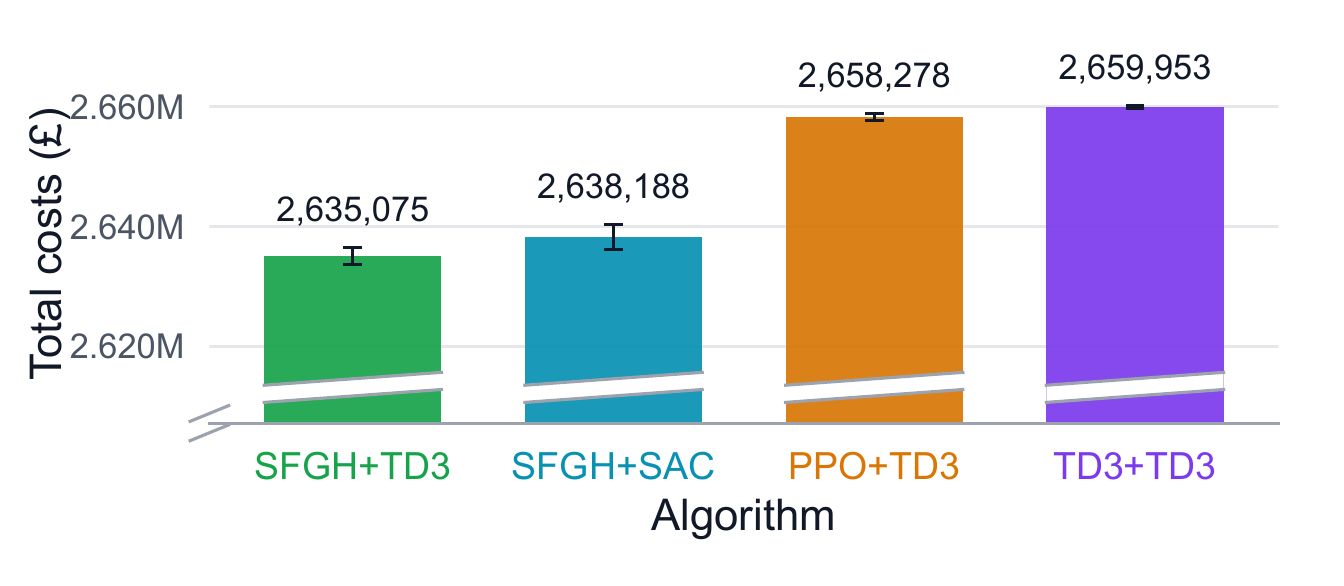}
    \caption{Total annual cost.}
    \label{fig:framework_cost}
\end{subfigure}

\caption{Test-average performance of the four scheduler--dispatcher combinations.}
\label{fig:framework_comparison}
\end{figure*}

PPO+TD3 and TD3+TD3 use three-stage training: first training the vehicle scheduler with a fixed rule-based dispatcher, then training the TD3 dispatcher with the learned scheduler fixed, and finally updating both policies under their respective objectives. Stage transitions are triggered by reward plateaus: consecutive 100-epoch reward averages are compared every 50 epochs, and a transition occurs after four consecutive checks without an improvement bigger than 0.01. The maximum training budget for this procedure is 18,000 epochs, with early stopping allowed in the final joint-training stage. Each of the four combinations was independently trained using five seeds, 0, 1, 21, 42, and 123, and evaluated on the same three year-long test datasets. Table~\ref{tab:framework_comparison} reports the mean and standard deviation across training runs for each dataset. Figure~\ref{fig:framework_comparison} presents these test-average results.

\begin{table*}[htbp]
\centering
\begingroup
\color{black}
\arrayrulecolor{black}
\captionsetup{labelfont={color=black},textfont={color=black}}

\caption{Performance comparison of four scheduler--dispatcher combinations across five independent training runs.}
\label{tab:framework_comparison}
\setlength{\tabcolsep}{3pt}
\renewcommand{\arraystretch}{1.1}
\small

\resizebox{\textwidth}{!}{%
\begin{tabular}{llccccccc}
\toprule
\multirow{2}{*}{Dataset} &
\multirow{2}{*}{Algorithm} &
\multicolumn{3}{c}{Service delays (h)} &
\multicolumn{4}{c}{Costs (\pounds)} \\
\cmidrule(lr){3-5}
\cmidrule(lr){6-9}
& & Total & Wait & Overflow & Total & Electricity & Hydrogen & PV \\
\midrule

\multirow{4}{*}{Test 1}
& SFGH+TD3
& $8791\pm0$ & $7762\pm0$ & $1029\pm0$
& $2625227\pm1258$ & $217445\pm238$ & $2403628\pm1392$ & $4154\pm361$ \\
& SFGH+SAC
& $8791\pm0$ & $7762\pm0$ & $1029\pm0$
& $2628167\pm1994$ & $217040\pm1876$ & $2401831\pm1001$ & $9296\pm417$ \\
& PPO+TD3
& $8921\pm7$ & $7898\pm8$ & $1022\pm4$
& $2648907\pm518$ & $219681\pm62$ & $2417138\pm371$ & $12088\pm117$ \\
& TD3+TD3
& $9148\pm33$ & $7899\pm7$ & $1249\pm36$
& $2650860\pm348$ & $219900\pm49$ & $2418606\pm252$ & $12354\pm86$ \\
\midrule

\multirow{4}{*}{Test 2}
& SFGH+TD3
& $8883\pm0$ & $7938\pm0$ & $945\pm0$
& $2665727\pm1415$ & $213762\pm245$ & $2447576\pm1556$ & $4389\pm335$ \\
& SFGH+SAC
& $8883\pm0$ & $7938\pm0$ & $945\pm0$
& $2669800\pm2166$ & $215444\pm2010$ & $2445182\pm1057$ & $9174\pm442$ \\
& PPO+TD3
& $8997\pm15$ & $8059\pm12$ & $938\pm4$
& $2689000\pm511$ & $215753\pm88$ & $2460897\pm359$ & $12350\pm158$ \\
& TD3+TD3
& $9235\pm30$ & $8072\pm6$ & $1163\pm32$
& $2689606\pm157$ & $215925\pm32$ & $2461227\pm130$ & $12454\pm38$ \\
\midrule

\multirow{4}{*}{Test 3}
& SFGH+TD3
& $8441\pm0$ & $7549\pm0$ & $892\pm0$
& $2614271\pm1493$ & $215457\pm205$ & $2395296\pm1477$ & $3518\pm214$ \\
& SFGH+SAC
& $8441\pm0$ & $7549\pm0$ & $892\pm0$
& $2616597\pm2109$ & $214795\pm1969$ & $2392957\pm1008$ & $8845\pm529$ \\
& PPO+TD3
& $8571\pm13$ & $7680\pm10$ & $891\pm3$
& $2636928\pm660$ & $217043\pm115$ & $2407996\pm423$ & $11890\pm203$ \\
& TD3+TD3
& $8783\pm40$ & $7687\pm9$ & $1096\pm37$
& $2639392\pm254$ & $217397\pm53$ & $2409955\pm227$ & $12040\pm59$ \\
\midrule

\multirow{4}{*}{Test Avg.}
& SFGH+TD3
& $8705\pm0$ & $7750\pm0$ & $955\pm0$
& $2635075\pm1389$ & $215555\pm229$ & $2415500\pm1475$ & $4020\pm303$ \\
& SFGH+SAC
& $8705\pm0$ & $7750\pm0$ & $955\pm0$
& $2638188\pm2090$ & $215760\pm1952$ & $2413323\pm1022$ & $9105\pm463$ \\
& PPO+TD3
& $8830\pm12$ & $7879\pm10$ & $950\pm4$
& $2658278\pm563$ & $217492\pm88$ & $2428677\pm384$ & $12109\pm159$ \\
& TD3+TD3
& $9055\pm34$ & $7886\pm7$ & $1169\pm35$
& $2659953\pm253$ & $217741\pm45$ & $2429929\pm203$ & $12283\pm61$ \\
\bottomrule
\end{tabular}%
}

\vspace{3pt}
\begin{minipage}{\textwidth}
\footnotesize
\color{black}
\textit{Notes:} PV denotes the curtailment opportunity cost. All values are rounded to integers.
\end{minipage}

\arrayrulecolor{black}
\endgroup
\end{table*}

SFGH achieves lower mean total service delay than both learned schedulers on all three test datasets. Its test-average delay is 8,705 h, compared with 8,830 h for PPO+TD3 and 9,055 h for TD3+TD3. The identical delays and zero run-to-run scheduling variation of the two SFGH-based combinations reflect the deterministic scheduler's independence from downstream dispatch. Together with Table~\ref{tab:1}, these results demonstrate SFGH's scheduling advantage over both the evaluated heuristic approach and the added DRL-based optimisation methods. SFGH+TD3 also achieves the lowest annual cost on every test dataset, providing favourable service and cost performance without requiring scheduler training.

Under the same SFGH schedule, TD3 achieves smaller cost standard deviations and lower mean annual costs than SAC on all three test datasets. The corresponding average standard deviations are \pounds1,389 and \pounds2,090, respectively, indicating more stable annual cost performance for TD3 across independent training runs. 
% TD3's cost advantage mainly arises from lower PV-curtailment costs. 
Although PPO+TD3 and TD3+TD3 exhibit smaller cost standard deviations than the two SFGH-based combinations, both have higher mean costs and service delays than SFGH+TD3. Their lower dispersion therefore indicates more consistent cost outcomes across training runs, rather than better service or economic performance. These results support the selection of SFGH+TD3 based on its favourable scheduling and cost performance.

Let $N_t^{wt}=|S_t^{wt,ev}|+|S_t^{wt,hv}|$, $N_t^{cand}=|S_t^{ev}|+|S_t^{hv}|$, and $N^{pile}=N^{ec}+N^{hr}$ denote the total numbers of waiting vehicles, retained candidates, and service facilities, respectively. For a direct implementation using comparison sorting and repeated scans of candidate--pile pairs, candidate selection requires $O(N_t^{wt}\log(N_t^{wt}+1))$ operations. Each assignment iteration checks at most $N_t^{cand}N^{pile}$ compatible pairs, and at most $N_t^{cand}$ assignments are accepted from the candidate set selected for that interval. Assuming constant-time feasibility and marginal-benefit evaluation for each pair, the assignment stage has an upper bound of $O(N^{pile}(N_t^{cand})^2)$. Including demand aggregation, the worst-case per-interval time complexity is therefore $O(N_t^{wt}\log(N_t^{wt}+1)+N^{pile}(N_t^{cand})^2+N^{pile})$. Restricting the candidate set controls the repeated assignment workload, while candidate selection still depends on the full waiting pool.

Decision times were also measured and averaged over the one-year evaluation horizon, excluding model loading, simulation advancement, and environment-state updates. SFGH requires 0.128 ms per scheduling decision, compared with 1.360 ms for PPO and 1.336 ms for TD3. Energy-dispatch decisions require 0.327 ms for TD3 and 0.381 ms for SAC. These averages are well below the 15-min control interval, supporting online computational feasibility for the tested configuration. The recorded training times are approximately 22.6 h, 27.3 h, 68.0 h, and 81.7 h for SFGH+TD3, SFGH+SAC, PPO+TD3, and TD3+TD3, respectively. Thus, SFGH+TD3 combines favourable service and cost performance with lower scheduling overhead and avoids the additional training required by learned vehicle schedulers.

% Insert Table~\ref{tab:framework_comparison} here.
% Insert Figure~\ref{fig:framework_comparison} here.

\subsection{Generalisation under Varying Transport Demand and PV Penetration} \label{sec:4.5}
The preceding comparative and ablation studies were conducted under relatively high vehicle-arrival rates and individual service demands. To examine the generalisation capability of the learned dispatch policies beyond this baseline setting, six additional scenarios were evaluated along two dimensions: transport demand and PV penetration. The former examines operation under lighter transport loads, whereas the latter examines changes in renewable-generation availability under fixed transport conditions. The previously trained DRL policies were directly evaluated without retraining, and the rule-based and No-ESS strategies were evaluated under the corresponding scenario settings.

\textit{1) Variations in Transport Demand:}
Three reduced-demand scenarios were constructed by modifying the vehicle-arrival and service-demand models in \ref{app:1}. In Case 1, the EV and HV arrival scales were reduced from 5 to 4 and from 3 to 2.5, respectively; In Case 2, the EV baseline demand was reduced from 25 to 20 kWh and the HV baseline demand from 180 to 150 Nm$^3$; Case 3 combines both reductions. 
% These three datasets were generated using the same random seed of 123. The PV capacity, storage parameters, and charging/refuelling infrastructure were unchanged, and the meteorological and electricity-price inputs were taken from the same 2025 records used for testing.

\begin{table*}[t]
\centering
\caption{Results of different vehicle scheduling algorithms under reduced transport demand}
\label{tab:4}
\setlength{\tabcolsep}{10pt}
\renewcommand{\arraystretch}{0.9}
\resizebox{0.8\textwidth}{!}{%
\begin{tabular}{lllllllll}
\toprule
\multicolumn{1}{c}{\multirow{2}{*}{Dataset}} &
\multicolumn{1}{c}{\multirow{2}{*}{Algorithm}} &
\multicolumn{3}{c}{All vehicles (h)} &
\multicolumn{2}{c}{EV (h)} &
\multicolumn{2}{c}{HV (h)} \\
\cmidrule(lr){3-5}
\cmidrule(lr){6-7}
\cmidrule(lr){8-9}
& &
Total & Wait & Overflow &
Wait & Overflow &
Wait & Overflow \\
\midrule
Case 1 & SFGH      & 1252 & 1249 & 3  & 560  & 0   & 689  & 3  \\
       & FIFO + EF & 1309 & 1305 & 4  & 592  & 0   & 713  & 4  \\
       & DA + EF   & 1252 & 1249 & 3  & 560  & 0   & 689  & 3  \\
       & LS + EF   & 1387 & 1383 & 4  & 639  & 0   & 744  & 4  \\
       & SPT + EF  & 1249 & 1222 & 27 & 552  & 4   & 670  & 23 \\[4pt]
Case 2 & SFGH      & 4474 & 4414 & 60  & 2256 & 18  & 2158 & 42  \\
       & FIFO + EF & 4668 & 4605 & 63  & 2358 & 19  & 2247 & 44  \\
       & DA + EF   & 4474 & 4414 & 60  & 2256 & 18  & 2158 & 42  \\
       & LS + EF   & 4899 & 4833 & 66  & 2485 & 20  & 2348 & 46  \\
       & SPT + EF  & 4630 & 4169 & 461 & 2154 & 242 & 2015 & 219 \\[4pt]
Case 3 & SFGH      & 478 & 478 & 0 & 223 & 0 & 255 & 0 \\
       & FIFO + EF & 505 & 505 & 0 & 235 & 0 & 270 & 0 \\
       & DA + EF   & 478 & 478 & 0 & 223 & 0 & 255 & 0 \\
       & LS + EF   & 540 & 540 & 0 & 254 & 0 & 286 & 0 \\
       & SPT + EF  & 474 & 472 & 1 & 221 & 0 & 251 & 1 \\
\bottomrule
\end{tabular}%
}
\end{table*}
As shown in Table \ref{tab:4}, SFGH maintains good scheduling performance across the three reduced-demand scenarios, achieving lower total service delays than FIFO+EF and LS+EF in every case. 
SFGH matches DA+EF in total service delay across all three reduced-demand scenarios at the reported precision. 
SPT+EF has a slight advantage in total delay in Cases 1 and 3, outperforming SFGH by only 3 h and 4 h, respectively. However, SFGH produces less overflow: 3 h compared with 27 h in Case 1, and 0 h compared with 1 h in Case 3, indicating better control of excessive waiting in these scenarios. Overall, SFGH remains competitive across the reduced-demand scenarios, demonstrating its robust scheduling performance. Table \ref{tab:5} further shows that the TD3-based approach continues to achieve the lowest annual operational cost among all compared approaches in all three scenarios. Although lighter transport demand reduces absolute energy requirements and operating costs, the proposed TD3-based approach maintains its superior performance. 
% These results demonstrate that the proposed hierarchical framework maintains its effectiveness under substantially different demand conditions without additional training, indicating good generalisation capability and robustness to changes in transport demand.

\textit{2) Variations in PV Penetration:} To further assess generalisation under changes in renewable generation availability, additional experiments were conducted using Test 1 as the reference dataset. PV penetration is defined here as the ratio of PV capacity to the peak electrical demand \cite{yang2018data}. In addition to the baseline penetration of approximately 62\% reported above, Cases 4--6 were evaluated at 0\%, 100\%, and 150\%, representing no PV, PV capacity matching peak demand, and PV capacity exceeding peak demand, respectively. These scenarios were constructed by adjusting the panel area $A$ in (\ref{eq:2.1.1}), while retaining the real-world meteorological data and other settings of Test 1. Sufficient installation area was assumed for the expanded PV configurations. Under the directional scheduling-to-dispatch structure, changing PV penetration does not affect the vehicle scheduling results. Therefore, only the energy dispatch results for Cases 4--6 are added to Table~\ref{tab:5}, with all dispatch methods using the same SFGH scheduler. 

As shown in Table~\ref{tab:5}, TD3 achieves the lowest total annual cost across Cases 4--6, demonstrating that the learned dispatch policy remains effective both without PV generation and with increased PV capacity. At 100\% and 150\% penetration, TD3 also achieves the lowest PV-curtailment cost among the evaluated approaches, indicating effective management of surplus renewable energy. 
The slight increase in total cost from 100\% to 150\% penetration occurs because the increase in PV curtailment opportunity costs outweighs the reductions in electricity and hydrogen procurement costs. This finding suggests that higher PV capacity does not necessarily lead to lower total costs; instead, PV capacity should be appropriately matched to the characteristics of the application scenario. Together with Cases 1--3, these results show that the previously trained policy retains its comparative cost advantage under changes in both transport demand and PV capacity, supporting the generalisation capability of the proposed framework.

\begin{table*}[t]
\centering
\caption{Annual costs of different approaches for energy dispatch layer under transport-demand variations and different PV penetration levels}
\label{tab:5}
\setlength{\tabcolsep}{3.5pt}
\renewcommand{\arraystretch}{1.05}
\small
\resizebox{\textwidth}{!}{%
\begin{tabular}{llrrrrrrrr}
\toprule
\multirow{2}{*}{Scenario} &
\multirow{2}{*}{Cost item (\pounds)} &
\multicolumn{8}{c}{Approach} \\
\cmidrule(lr){3-10}
& & TD3 & SAC & DDPG & PPO & DQN & DDQN & Rule-based & No-ESS \\
\midrule

\multirow{4}{*}{Case 1}
& Total       & 1544599 & 1556521 & 1546523 & 1570677 & 1567349 & 1555030 & 1599035 & 1714670 \\
& Electricity & 125616  & 124274  & 126525  & 128773  & 133146  & 129767  & 129146  & 152785  \\
& Hydrogen    & 1411056 & 1414041 & 1411744 & 1433657 & 1417580 & 1411687 & 1444097 & 1523494 \\
& PV          & 7927    & 18206   & 8254    & 8247    & 16623   & 13576   & 25792   & 38391   \\[3pt]

\multirow{4}{*}{Case 2}
& Total       & 1964543 & 1973922 & 1966249 & 1996508 & 1983488 & 1971239 & 2015222 & 2127544 \\
& Electricity & 173600  & 172904  & 174656  & 182902  & 182208  & 177462  & 179448  & 202274  \\
& Hydrogen    & 1785678 & 1788581 & 1786154 & 1807673 & 1789551 & 1784356 & 1817608 & 1896070 \\
& PV          & 5265    & 12437   & 5439    & 5933    & 11729   & 9421    & 18166   & 29200   \\[3pt]

\multirow{4}{*}{Case 3}
& Total       & 1144682 & 1156001 & 1148473 & 1167995 & 1170151 & 1158991 & 1195056 & 1312508 \\
& Electricity & 99801   & 97966   & 100648  & 99098   & 106818  & 104166  & 102291  & 124979  \\
& Hydrogen    & 1034439 & 1035706 & 1037000 & 1058015 & 1042051 & 1037537 & 1063835 & 1143015 \\
& PV          & 10442   & 22329   & 10825   & 10882   & 21282   & 17288   & 28930   & 44514   \\

% \midrule
% \multicolumn{10}{l}{\textit{B. PV-capacity variations with fixed Test 1 transport and price inputs}} \\
% \midrule

\multirow{4}{*}{\revblue{Case 4}}
& \revblue{Total}
& \revblue{2713837} & \revblue{2714120} & \revblue{2723735} & \revblue{2785316}
& \revblue{2724886} & \revblue{2719376} & \revblue{2743646} & \revblue{2832973} \\
& \revblue{Electricity}
& \revblue{293257} & \revblue{291818} & \revblue{294031} & \revblue{320510}
& \revblue{304615} & \revblue{298121} & \revblue{306248} & \revblue{321001} \\
& \revblue{Hydrogen}
& \revblue{2420580} & \revblue{2422302} & \revblue{2429704} & \revblue{2464806}
& \revblue{2420271} & \revblue{2421255} & \revblue{2437398} & \revblue{2511972} \\
& \revblue{PV}
& \revblue{0} & \revblue{0} & \revblue{0} & \revblue{0}
& \revblue{0} & \revblue{0} & \revblue{0} & \revblue{0} \\[3pt]

\multirow{4}{*}{\revblue{Case 5}}
& \revblue{Total}
& \revblue{2595823} & \revblue{2619654} & \revblue{2601626} & \revblue{2635485}
& \revblue{2615724} & \revblue{2615061} & \revblue{2668484} & \revblue{2792017} \\
& \revblue{Electricity}
& \revblue{192491} & \revblue{192341} & \revblue{193455} & \revblue{191341}
& \revblue{199549} & \revblue{194293} & \revblue{198851} & \revblue{221905} \\
& \revblue{Hydrogen}
& \revblue{2389040} & \revblue{2394868} & \revblue{2391204} & \revblue{2425363}
& \revblue{2392850} & \revblue{2395961} & \revblue{2431515} & \revblue{2511972} \\
& \revblue{PV}
& \revblue{14293} & \revblue{32446} & \revblue{16968} & \revblue{18781}
& \revblue{23326} & \revblue{24807} & \revblue{38118} & \revblue{58140} \\[3pt]

\multirow{4}{*}{\revblue{Case 6}}
& \revblue{Total}
& \revblue{2604231} & \revblue{2644806} & \revblue{2617423} & \revblue{2632819}
& \revblue{2622966} & \revblue{2626261} & \revblue{2674647} & \revblue{2836228} \\
& \revblue{Electricity}
& \revblue{175146} & \revblue{174963} & \revblue{175615} & \revblue{163994}
& \revblue{177783} & \revblue{177038} & \revblue{181372} & \revblue{204702} \\
& \revblue{Hydrogen}
& \revblue{2375137} & \revblue{2387644} & \revblue{2376671} & \revblue{2413530}
& \revblue{2374358} & \revblue{2383564} & \revblue{2420542} & \revblue{2511972} \\
& \revblue{PV}
& \revblue{53948} & \revblue{82200} & \revblue{65137} & \revblue{55295}
& \revblue{70826} & \revblue{65660} & \revblue{72733} & \revblue{119555} \\

\bottomrule
\end{tabular}%
}
\vspace{2pt}
\begin{minipage}{\textwidth}
\footnotesize
Note: Cases 1--3 represent reduced vehicle arrivals, reduced individual service demand, and both reductions, respectively. Cases 4--6 correspond to PV penetration levels of 0\%, 100\%, and 150\%, respectively.
\end{minipage}
\end{table*}

% Compared with the original testing environment, the reduced-demand scenarios result in lower charging and refuelling requirements, thereby changing the utilisation patterns of the battery, hydrogen tank, and renewable-energy resources. The objective of this experiment is to examine whether the dispatch policies learned under high-demand conditions remain effective when deployed under substantially lighter transport loads.

\subsection{Discussion}\label{sec:4.6}

\textbf{1) Practical Implementation of Vehicle Scheduling:}
From a practical perspective, the proposed scheduling strategy is designed from the viewpoint of the EHTS operator rather than individual vehicle users. By jointly considering waiting time, overflow time, and energy demand, the scheduler seeks to minimise the overall service delay of the EHTS instead of strictly preserving a first-in-first-out service order. Nevertheless, this does not imply that drivers are required to physically change their positions in a queue. In practice, the proposed scheduling strategy can be implemented through operational mechanisms that are already adopted or increasingly available in modern charging infrastructures, such as reservation systems, digital queue management, designated waiting areas, bay assignment upon arrival, or automated parking services. Under these implementations, vehicles can be directed to appropriate chargers or hydrogen refuellers immediately after registration, enabling flexible resource allocation while maintaining a practical user experience.

The benefit of WISA depends on the decision-interval length relative to vehicle service durations. With shorter intervals, a scheduler that assigns vehicles only at decision boundaries can reconsider assignments more frequently, reducing the potential idle time between service completion and the next decision instant. The additional benefit of WISA may therefore become smaller. Longer intervals provide greater scope for unused service time to accumulate when within-interval reassignment is prohibited. However, the resulting delay reduction also depends on arrival intensity, service durations, and the availability of waiting vehicles; its magnitude at other temporal resolutions cannot be established from the present 15-min experiments alone. An alternative is event-driven reassignment, in which a new scheduling decision is triggered whenever a charger or refueller becomes available. The present WISA mechanism captures within-interval pile reuse by planning successive services at the interval boundary, while retaining a fixed decision interval for energy dispatch. A fully event-driven scheduler could additionally respond to updated queue information and arrivals between decision boundaries. Such an extension would require corresponding updates to the charging and refuelling demands supplied to the energy-dispatch layer.

\textbf{2) Sequential Architecture and Modularity:} The proposed framework follows a service-first, demand-driven structure: vehicle scheduling determines charging and refuelling assignments, and energy dispatch subsequently optimises supply to meet the resulting demands. Deriving energy requirements directly from vehicle-service decisions maintains consistency between scheduled services and downstream demand. The modular structure allows either scheduling or dispatch methods to be replaced without redesigning the overall architecture, as illustrated by the alternative scheduler--dispatcher combinations in Section~\ref{sec:framework_comparison}. It therefore preserves the operational dependency between the layers while providing flexibility in algorithm selection and implementation.

Energy-side feedback offers a potential extension to this sequential structure. The current SFGH rules do not use energy prices, PV availability, storage SOCs, or dispatch-related rewards, so returning these signals alone would not change scheduling decisions. Incorporating such feedback would require extending the scheduling formulation to consider both service quality and energy costs. Although PPO+TD3 and TD3+TD3 in Section~\ref{sec:framework_comparison} update both policies during their final training stage, they retain separate objectives and unidirectional information flow. Explicit energy-side feedback to vehicle scheduling could therefore offer a promising direction for further exploration, with its potential benefits assessed in terms of service quality and operational costs.

\textbf{3) Model Assumptions and Extensions:} Pile heterogeneity is another consideration, as different charger powers and refueller flow rates would affect service completion times and the opportunities for within-interval pile reuse. Although the present model assumes identical piles within each service type, the SFGH assignment procedure evaluates individual vehicle–pile pairs and tracks pile availability. It could therefore be extended by using pile-specific service rates to calculate the service duration and completion time of each feasible assignment, with vehicle–pile compatibility included where necessary. These estimates would then be used in the greedy assignment evaluation and subsequent availability updates. The charging and refuelling demands passed to energy dispatch would also need to be calculated using each pile’s service rate and occupied duration. This extension would retain the sequential assignment structure, but its scheduling performance under heterogeneous configurations remains to be evaluated.

The electrolyser model adopts constant specific electricity consumption and does not explicitly represent part-load performance, electrolyser-specific operating limits, or startup/shutdown dynamics. Although electrolysis is constrained by surplus PV availability and hydrogen-tank charging limits, these constraints do not fully capture the electrolyser’s operating characteristics. Load-dependent electricity consumption could change hydrogen production for a given PV allocation, thereby affecting pipeline hydrogen purchases and operating costs. Minimum operating loads and startup/shutdown requirements could also restrict the utilisation of intermittent surplus PV, potentially increasing curtailment. The reported economic results should therefore be interpreted under these modelling assumptions. Incorporating a more detailed electrolyser model would allow these effects on dispatch decisions and economic performance to be assessed.

\textbf{4) Scalability and Adaptation:} The adaptability and computational requirements of the proposed framework should also be considered across different EHTS configurations. For the vehicle-scheduling layer, the scheduling procedure can be adapted to different station configurations by updating the pile states and candidate-subset sizes. In this work, the candidate subsets are empirically configured as approximately twice the numbers of chargers and hydrogen refuellers. This setting limits the number of vehicles considered during assignment evaluation, although candidate selection still depends on the waiting pool. The ratio is a practical configuration rather than an algorithmic requirement and can be adjusted for other station scales. The complexity analysis in Section~\ref{sec:framework_comparison} makes explicit the effects of waiting-pool size, candidate-set size, and facility count. Limiting the candidate set controls repeated assignment evaluation, although enlarging both the candidate set and the infrastructure increases this workload. The measured execution times support online operation for the evaluated configuration, while runtime at larger scales should be assessed against the corresponding decision-time requirements.

For the energy-dispatch layer, the controller operates on aggregated energy states rather than individual vehicle information, allowing system parameters to be updated while retaining the hierarchical architecture. The experiments in Section~\ref{sec:4.5} demonstrate that the trained policy maintains competitive performance under the evaluated transport-demand and PV-penetration variations without retraining. Adaptation to other changes, such as different storage capacities or device operating characteristics, may require further validation and, where necessary, policy retraining. When retraining is required, it can be performed offline, while online deployment remains computationally efficient, as each dispatch decision requires only a single forward pass through the trained neural network.

% the solver-free scheduling algorithm avoids repeatedly solving mathematical optimisation problems at every decision interval, making the framework particularly suitable for real-time operation in scalable EHTSs. Moreover, increasing the number of waiting vehicles does not proportionally increase the computational burden because scheduling decisions are generated based on bounded candidate subsets rather than the entire waiting pool. In this work, the candidate subset sizes are empirically configured as approximately twice the numbers of chargers and hydrogen refuellers, so that the scheduling effort scales with the charging/refuelling infrastructure rather than the instantaneous number of waiting vehicles. The specific ratio serves as a practical configuration rather than an algorithmic requirement and can be proportionally adapted for different station scales. When the charging/refuelling infrastructure itself is expanded, additional pile states are introduced to represent the availability of chargers and hydrogen refuellers, but the scheduling procedure itself remains unchanged.

\section{Conclusion} \label{sec:5}
This paper proposed a hierarchical optimisation framework for integrated electric–hydrogen–transport systems that makes explicit the dependency between station-level vehicle scheduling and downstream multi-energy dispatch. Rather than treating charging and refuelling demand as an externally given profile, the framework generates it through real-time scheduling of heterogeneous EVs and HVs and propagates it to the dispatch layer across the coupled electricity–hydrogen networks. The framework integrates the SFGH algorithm for real-time EV/HV service scheduling with a TD3 approach for battery-storage operation, hydrogen-tank operation, and PV generation allocation. Simulation results showed that the proposed SFGH achieved the lowest test-average total service delay in the baseline scenario and maintained competitive performance across the reduced-demand scenarios. The TD3-based approach delivered the lowest total annual cost among the evaluated dispatch methods under both the baseline conditions and the additional transport-demand and PV-penetration scenarios, reflecting its overall performance across electricity, hydrogen, and PV-curtailment cost components. Together with the comparative and ablation results, these findings support the effectiveness and generalisation capability of the hierarchical framework. Future work will explore coordinated learning between vehicle scheduling and energy dispatch by feeding energy-side states and learning signals back to the scheduling layer, and assess whether this closer interaction offers advantages over the current sequential decision-making structure.

\section*{Acknowledgments}
This work was supported in part by the Clean Energy and Equitable Transportation Solutions (CLEETS) NSF–UKRI Global Centre under NSF Award No. 2330565 and UKRI Award No. EP/Y026233/1.

\appendix
\renewcommand{\thefigure}{A.\arabic{figure}}
\setcounter{figure}{0}
\setcounter{table}{0}

\section{Synthetic Dataset Generation} \label{app:1}
A hybrid dataset with a 15-minute temporal resolution is constructed to evaluate the proposed hierarchical optimisation framework under long-term operating conditions. The dataset combines real-world meteorological and electricity market data with synthetically generated hydrogen prices, vehicle arrivals, and service demands. 

\textit{1) Solar Irradiance and Ambient Temperature}

Solar irradiance and ambient temperature were derived from the Copernicus Climate Change Service (C3S) ERA5 reanalysis dataset \cite{era5_timeseries}. Hourly records of \textit{surface solar radiation downwards} ($G_t$) and \textit{2m temperature} ($T_t^{\mathrm{am}}$) for the years 2024 and 2025 were obtained for the South Wales region (51.50$^\circ$N, 3.25$^\circ$W). The resulting time series were subsequently upsampled to a 15-minute resolution through linear interpolation.

\textit{2) Electricity Price}

Half-hourly wholesale electricity prices for 2024--2025 were obtained from the GB Market Index Price (MIP) published by Elexon BMRS \cite{elexon_mip}. The MIP represents the wholesale electricity price in the Great Britain market and provides a representative benchmark for short-term electricity trading. However, the electricity procurement cost faced by commercial charging operators is not solely determined by wholesale market prices, but also includes transmission and distribution network charges, balancing costs, supplier operating costs, and profit margins \cite{uk_bill_components}. Accordingly, a simplified surcharge factor of 30\% was applied to the wholesale electricity price as a representative approximation of these additional procurement-related costs. Since the focus of this study is system operational optimisation rather than electricity market risk assessment, prices above the 99th percentile were clipped to reduce the influence of rare extreme price spikes. Moreover, although negative wholesale electricity prices occasionally occur in the GB market due to periods of excess renewable generation and low demand, the effective procurement cost faced by charging operators is rarely negative after accounting for the additional charges described above. Therefore, a minimum electricity price of £0.01/kWh was imposed to ensure non-negative procurement costs while preserving low-price market signals. Finally, the half-hourly electricity price series was upsampled to a 15-minute resolution through linear interpolation, yielding the electricity procurement price $p_t^{elc}$ used in the EHTS simulation framework.

\textit{3) Hydrogen Price}

As the UK does not yet have a mature hydrogen market with widely available time-varying pricing data, a synthetic hydrogen price model is built up to represent plausible temporal variations in hydrogen procurement costs. The model is constructed by combining a baseline procurement price with several temporal adjustment components, including seasonal effects, weekly demand fluctuations, intra-day variations, stochastic market uncertainty, and an electricity-price coupling term \cite{parra2019review}.

The baseline procurement price is derived from the renewable hydrogen production cost reported by the European Hydrogen Observatory for the United Kingdom in 2024, which is 3.4 \euro/kg H$_2$ \cite{eho_hydrogen_cost_2024}. This corresponds to approximately 0.26 $\mathrm{\pounds/Nm^{3}}$, assuming an exchange rate of 0.85~\pounds/\euro{} and a conversion factor of 1 kg H$_2$ = 11.126 Nm$^3$. Similar to the treatment applied to electricity prices, an additional 30\% surcharge is introduced to account for pipeline transport, compression, distribution losses, supplier operating costs, and other procurement-related expenses. Consequently, the baseline procurement price is set to $p_0^{h2}=0.34 \mathrm{\pounds/Nm^{3}}$. Based on this baseline, the time-varying hydrogen procurement price is defined as
\begin{equation}
p_t^{h2} = p_0^{h2} \left( 1 + s_t^{h2} + w_t^{h2} + c_t^{h2}
+ d_t^{h2} + \eta_t^{h2} \right),
\label{eq:h2_price}
\end{equation}
where $s_t^{h2}$ and $w_t^{h2}$ represent seasonal and weekly variations, $c_t^{h2}$ denotes the electricity-price coupling effect, $d_t^{h2}$ captures intra-day fluctuations, and $\eta_t^{h2}$ represents stochastic market uncertainty. The seasonal and weekly components are modelled as
\begin{equation}
s_t^{h2} = \alpha_s^{h2} \cos \left( 2\pi \frac{\mathcal{D}_t}{365}\right),
\qquad
w_t^{h2} = \alpha_w^{h2} \cos(2\pi\frac{ \mathcal{W}_t}{7}),
\label{eq:h2_season}
\end{equation}
where $\alpha_s^{h2}$ and $\alpha_w^{h2}$ control the magnitudes of the seasonal and weekly price variations, respectively; and $\mathcal{D}_t$ and $\mathcal{W}_t$ denote the day-of-year and day-of-week indices, respectively. To account for the dependence of hydrogen production costs on electricity prices, an electricity-coupling component is introduced:
\begin{equation}
c_t^{h2}
=
k_c^{h2}
\left(
\frac{p_t^{elc}}{\bar{p}^{elc}}
-1
\right),
\label{eq:h2_coupling}
\end{equation}
where $k_c^{h2}$ represents the coupling coefficient and $\bar{p}^{elc}$ denotes the average electricity price over the study period. The intra-day fluctuation term is represented by combined daily and half-daily oscillations:
\begin{equation}
d_t^{h2}
=
\alpha_{24}^{h2}\gamma_{24,t}
\sin
\left(
\frac{2\pi h_t}{24}
+
\phi_{24,t}
\right)
+
\alpha_{12}^{h2}\gamma_{12,t}
\sin
\left(
\frac{2\pi h_t}{12}
+
\phi_{12,t}
\right),
\label{eq:h2_intraday}
\end{equation}
where $\alpha_{24}^{h2}$ and $\alpha_{12}^{h2}$ are the amplitudes of the 24-hour and 12-hour oscillations, respectively. The terms $\gamma_{24,t}$ and $\gamma_{12,t}$ denote random amplitude modifiers, while $\phi_{24,t}$ and $\phi_{12,t}$ denote random phase shifts. The 24-hour component captures the primary daily variation, while the 12-hour component allows for secondary semi-daily fluctuations, such as morning and evening demand effects. Random amplitude modifiers and phase shifts are introduced to avoid identical daily patterns and to generate more realistic temporal variability. To further represent short-term stochastic market uncertainty, a temporally correlated noise term is introduced:
\begin{equation}
\eta_t^{h2}
=
(1-\alpha_{\eta}^{h2})\eta_{t-1}^{h2}
+
\alpha_{\eta}^{h2}\nu_t,
\qquad
\nu_t \sim \mathcal{N}(0,\sigma_{\eta}^2),
\label{eq:h2_noise}
\end{equation}
where $\sigma_{\eta}$ controls the magnitude of stochastic noise and $\alpha_{\eta}^{h2}$ controls the temporal correlation level. This formulation avoids unrealistically independent fluctuations between adjacent time steps and produces smoother short-term hydrogen price variations. 
% Finally, the hydrogen price is constrained to a realistic commercial range:
% \begin{equation}
% p_t^{h2}
% =
% \mathrm{clip}
% \left(
% p_t^{h2},
% p_{\min}^{h2},
% p_{\max}^{h2}
% \right).
% \label{eq:h2_clip}
% \end{equation}
The parameter values used in this study are summarised in Table~\ref{tab:app_h2}. 
\begin{table}[!htbp]
\centering
\footnotesize
\setlength{\tabcolsep}{4pt}
\caption{Parameters used for synthetic hydrogen price generation.}
\label{tab:app_h2}
\begin{tabular}{llllll}
\hline
Parameter & Value & Description &
Parameter & Value & Description \\
\hline

$p_0^{h2}$
& 0.34
& Baseline procurement price
&
$\alpha_s^{h2}$
& 0.08
& Seasonal amplitude
\\

$\alpha_w^{h2}$
& 0.03
& Weekly amplitude
&
$k_c^{h2}$
& 0.08
& Electricity coupling coefficient
\\

$\alpha_{24}^{h2}$
& 0.04
& Daily oscillation amplitude
&
$\alpha_{12}^{h2}$
& 0.02
& Half-daily oscillation amplitude
\\

$\sigma_{\eta}$
& 0.015
& Noise standard deviation
&
$\alpha_{\eta}^{h2}$
& 0.15
& Noise correlation coefficient
\\

$\gamma_{24,t},\gamma_{12,t}$
& $U(0.8,1.2)$
& Random amplitude modifiers
&
$\phi_{24,t},\phi_{12,t}$
& $U(0,2\pi)$
& Random phase shifts
\\

\hline
\end{tabular}
\end{table}

\textit{4) EV and HV Arrival Processes}

EV and HV arrivals are modelled as non-homogeneous Poisson processes with time-varying arrival intensities \cite{traffic_poisson_arrivals,kurtz2020predicting}. The raw arrival intensities are constructed from baseline arrival scales together with time-of-day, weekday/weekend, and stochastic traffic variations. Specifically, the baseline arrival scales are set to 5 and 3 for EVs and HVs, respectively. The influence of different arrival-scale settings is further investigated in Section \ref{sec:4.5} through additional demand scenarios. To capture diurnal traffic variations, two arrival-rate regimes are considered. During the daytime period (09:00--21:00), the baseline arrival scales are multiplied by 0.6 for EVs and 0.5 for HVs. During the nighttime period (21:00--09:00), the corresponding multipliers become 0.2 and 0.25, respectively. To represent systematic differences between weekdays and weekends, EV/HV arrivals are further multiplied by 1.2 on weekdays and 0.8 on weekends, respectively. In addition, a mild stochastic perturbation sampled from $U(0.95,1.05)$ is introduced at every simulation step. The resulting raw arrival intensity is denoted by $\lambda_{t,\mathrm{raw}}^{x}$, where $x\in\{ev,hv\}$. To avoid unrealistically abrupt fluctuations and to represent traffic inertia, the raw arrival intensities are smoothed using an exponential moving average (EMA), i.e.,
\begin{equation} 
\lambda_t^{x} = (1-\alpha_{\lambda}) \lambda_{t-1}^{x} + \alpha_{\lambda} \lambda_{t,\mathrm{raw}}^{x}, \qquad x\in\{ev,hv\},
\label{eq:arrival_smoothing} 
\end{equation} 
where $\lambda_t^{x}$ denotes the smoothed arrival intensities and the smoothing factor $\alpha_{\lambda}$ is defined as
\begin{equation} 
\alpha_{\lambda} = \frac{\Delta t} {\tau_{\lambda}+\Delta t}, 
\label{eq:arrival_alpha} 
\end{equation} 
with a smoothing time constant of $\tau_{\lambda}=1$~h (4 simulation steps). To capture lower-demand days caused by weather conditions, holidays, or reduced fleet activity, a day-level traffic regime factor $\psi_d^{x}$ is introduced 
\begin{equation} 
\psi_d^{x} = \begin{cases} 0.6, & \text{low-demand day},\\ 1.0, & \text{otherwise}, \end{cases} \qquad x\in\{ev,hv\}, 
\label{eq:arrival_day_factor} 
\end{equation} 
where a day is classified as a low-demand day with probabilities of 20\% and 15\% for EVs and HVs, respectively. Finally, the numbers of EV and HV arrivals at simulation step $t$ are generated as 
\begin{equation} 
n_t^{ev} \sim \mathrm{Poisson} \left( \lambda_t^{ev}\psi_{d}^{ev} \right), \qquad n_t^{hv} \sim \mathrm{Poisson} \left( \lambda_t^{hv}\psi_{d}^{hv} \right), 
\label{eq:arrival_poisson} 
\end{equation} 

To prevent implausibly large arrival spikes, the generated arrivals are bounded by
\begin{equation} 
n_t^{ev} = \min(n_t^{ev},9), \qquad n_t^{hv} = \min(n_t^{hv},6). 
\label{eq:arrival_cap} 
\end{equation}

\textit{5) EV and HV Service Demand}

For each arriving vehicle, the charging or refuelling demand is generated according to the following stochastic demand model. For EVs, the demand-generation process starts from a daily baseline demand of 25 kWh and is increased by 5 kWh on weekdays. To represent day-to-day variability, a Gaussian perturbation with a standard deviation of 5 kWh is added. For HVs, the daily baseline hydrogen demand is set to 180 Nm$^3$ and is increased by 20 Nm$^3$ on weekdays. A Gaussian perturbation with a standard deviation of 20 Nm$^3$ is added to represent daily variability. These baseline demand settings are further varied in Section \ref{sec:4.5} to evaluate the generalisation capability of the proposed framework under different transport-demand conditions. EV demands are limited to 8--50 kWh per vehicle, while HV demands are limited to 50--350 Nm$^3$ per vehicle, thereby covering representative service demands ranging from private vehicles to commercial logistics vehicles. To capture systematic variations throughout the day, weak intra-day modulation is applied using sinusoidal profiles. EV demand exhibits a modulation amplitude of 8\%, while HV demand uses a smaller amplitude of 4\%. The modulation peaks occur during the afternoon period and represent typical variations in charging and refuelling behaviour over the course of a day. Additional short-term variability is introduced by applying a multiplicative perturbation independently sampled from $U(0.95,1.05)$ at each simulation step. The resulting time-dependent raw demand profile is denoted by $D_{t,\mathrm{raw}}^{x}$, which is then smoothed using an exponential moving average to represent temporal persistence in user behaviour
\begin{equation}
D_t^{x}=
(1-\alpha_D)
D_{t-1}^{x}
+
\alpha_D
D_{t,\mathrm{raw}}^{x},
\qquad
x\in\{ev,hv\},
\label{eq:demand_smoothing}
\end{equation}
where $D_t^{x}$ denotes the smoothed demand profile and the smoothing factor $\alpha_D$ is defined as
\begin{equation}
\alpha_D=
\frac{\Delta t}
{\tau_D+\Delta t},
\label{eq:demand_alpha}
\end{equation}
with $\tau_D=0.75$h (3 simulation steps). 

Most vehicles exhibit demand levels close to the smoothed time-step mean demand, while a small proportion of heavy-demand users is introduced to reflect heterogeneous charging and refuelling requirements. For EVs, the base vehicle-level demand multipliers are sampled from a normal distribution with a mean of 1.0 and a standard deviation of 0.07. Approximately 10\% of EVs are further classified as heavy-demand users and receive an additional multiplicative factor sampled uniformly from 1.15 to 1.35. For HVs, the base multipliers are sampled from a normal distribution with mean 1.0 and standard deviation 0.10, while approximately 15\% of HVs receive an additional multiplicative factor sampled uniformly from 1.2 to 1.6. The resulting multiplier $\eta_{k,t}^{x}$ is then used to determine the final demand of vehicle $k$ arriving at time step $t$
\begin{equation}
E_{k,t}^{x}=
D_t^{x}
\cdot
\eta_{k,t}^{x},
\qquad
x\in\{ev,hv\},
\label{eq:vehicle_demand}
\end{equation}

The above parameters governing vehicle arrival and service-demand profiles were empirically selected to generate representative operating conditions while maintaining realistic station utilisation throughout the simulation period.

\textit{6) Training and Testing Dataset Generation}

Using the data-generation procedures described above, one training dataset and three testing datasets are constructed. The training dataset is based on the 2024 solar irradiance, ambient temperature, and electricity price data, while hydrogen prices, EV/HV arrivals, and individual service demands are generated using a random seed of 42. For testing, the 2025 solar irradiance, ambient temperature, and electricity price data are employed. Three testing datasets are then completed using random seeds of 2024 (Test 1), 2025 (Test 2), and 2026 (Test 3), respectively, for the hydrogen price, EV/HV arrival, and vehicle demand models. Consequently, all testing datasets share identical meteorological conditions and electricity prices, while differing in hydrogen market fluctuations and transport demand patterns.

Different random seeds lead to different realisations of the underlying stochastic processes used to generate hydrogen prices, EV/HV arrivals, and vehicle-level service demands. Specifically, the seeds affect the stochastic disturbance terms, random amplitude modifiers, and random phase shifts in the hydrogen price model; the stochastic perturbations, low-demand-day realisations, and Poisson arrival processes in the EV/HV arrival model; and the Gaussian perturbations, short-term multiplicative perturbations, heavy-demand user assignments, and vehicle-level demand multipliers in the service-demand model. Consequently, the resulting test datasets exhibit very different hydrogen price trajectories, arrival patterns, individual charging/refuelling demands, and aggregate service demands despite sharing identical meteorological conditions and electricity prices. Evaluating the proposed framework across these datasets, therefore, provides a meaningful assessment of its robustness and generalisation capability under diverse yet realistic future operating scenarios.

\bibliographystyle{elsarticle-num} 
\bibliography{ehts}

@techreport{iea2025ev,
  author       = {{International Energy Agency}},
  title        = {Global EV Outlook 2025},
  institution  = {International Energy Agency},
  year         = {2025},
  note         = {Available at: \url{https://www.iea.org/reports/global-ev-outlook-2025}. Accessed: 2026-06-18}
}

@techreport{cpcatapult2025hgv,
  title        = {Zero Emission HGVs and Infrastructure: Infrastructure Planning and Delivery},
  institution  = {Connected Places Catapult},
  year         = {2025},
  note         = {Available at: \url{https://cp.catapult.org.uk/report/zero-emission-hgvs-and-infrastructure-infrastructure-planning-and-delivery/}. Accessed: 2026-06-18}
}

@misc{reuters_aegis2025,
  author       = {{Reuters}},
  title        = {British Startup Gets £100 Million Funding for Green Charging Stations},
  year         = {2025},
  note         = {Available at: \url{https://www.reuters.com/business/autos-transportation/british-startup-gets-122-million-funding-green-charging-stations-2025-01-20/}. Accessed: 2026-06-18}
}

@article{chen2025superconducting,
  title={Superconducting hydrogen-electricity multi-energy system for transportation hubs: Modeling, technical study and economic-environmental assessment},
  author={Chen, Yu and Chen, Xiaoyuan and Fu, Lin and Jiang, Shan and Shen, Boyang},
  journal={Applied Energy},
  volume={401},
  pages={126823},
  year={2025},
  publisher={Elsevier}
}

@article{salam2024charge,
  title={Charge scheduling optimization of electric vehicles: A comprehensive review of essentiality, perspectives, techniques, and security},
  author={Salam, Shereen Siddhara Abdul and Raj, Veena and Petra, Mohammad Iskandar and Azad, Abul Kalam and Mathew, Sathyajith and Sulthan, Sheik Mohammed},
  journal={IEEE Access},
  volume={12},
  pages={121010--121034},
  year={2024},
  publisher={IEEE}
}

@article{zhao2024two,
  title={A two-level charging scheduling method for public electric vehicle charging stations considering heterogeneous demand and nonlinear charging profile},
  author={Zhao, Zhonghao and Lee, Carman KM and Ren, Jingzheng},
  journal={Applied energy},
  volume={355},
  pages={122278},
  year={2024},
  publisher={Elsevier}
}

@article{tsaousoglou2023fair,
  title={Fair and scalable electric vehicle charging under electrical grid constraints},
  author={Tsaousoglou, Georgios and Giraldo, Juan S and Pinson, Pierre and Paterakis, Nikolaos G},
  journal={IEEE Transactions on Intelligent Transportation Systems},
  volume={24},
  number={12},
  pages={15169--15177},
  year={2023},
  publisher={IEEE}
}

@article{tan2026distributionally,
  title={Distributionally robust scheduling of electric-hydrogen integrated energy systems based on pipeline-road coordinated hydrogen transportation},
  author={Tan, Hong and Chen, Shun and Lin, Zhenjia and Wang, Qiujie and Mohamed, Mohamed A},
  journal={Applied Energy},
  volume={404},
  pages={127055},
  year={2026},
  publisher={Elsevier}
}

@article{li2026optimal,
  title={Optimal Chance-Constrained Scheduling of Hydrogen Storage System Based on Data-Driven Model Predictive Control in Microgrids},
  author={Li, Yonggang and Zhang, Andong and Zhang, Yuanjin and Su, Yaotong and Li, Longjiang and Pei, Errong},
  journal={IEEE Transactions on Industry Applications},
  year={2026},
  doi={10.1109/TIA.2026.3676495},
  publisher={IEEE}
}

@article{elsir2025holistic,
  title={A holistic risk-aware coordinated framework for coupled hydrogen transport, LOHC storage, and rich renewable grid optimization},
  author={Elsir, Mohamed and Al-Sumaiti, Ameena Saad},
  journal={Applied Energy},
  volume={392},
  pages={125979},
  year={2025},
  publisher={Elsevier}
}

@article{sun2025end,
  title={End-to-End Tri-Stage Low-Carbon Planning of Coupled Electricity--Hydrogen--Gas--Transportation Systems toward Hydrogen-Fueled Mobility},
  author={Sun, Yuxin and Lv, Yong and Wang, Guibin and Zhang, Xian and Zhang, Yan and Wan, Yanming and Wu, Yifeng and Liu, Chang},
  journal={IEEE Transactions on Industry Applications},
  volume={62},
  number={4},
  pages={5685--5698},
  year={2026},
  publisher={IEEE}
}

@article{yao2025holistic,
  title={A holistic power optimization approach for microgrid control based on deep reinforcement learning},
  author={Yao, Fulong and Zhao, Wanqing and Forshaw, Matthew and Song, Yang},
  journal={Neurocomputing},
  pages={131375},
  year={2025},
  publisher={Elsevier}
}

@article{jadidbonab2026drl,
  title={A DRL-Based Framework for Optimized Scheduling and Delivery in a Green Hydrogen Hub},
  author={Jadidbonab, Mohammad and Abdeltawab, Hussein and Mohamed, Yasser Abdel-Rady I},
  journal={IEEE Transactions on Industry Applications},
  year={2026},
  doi={10.1109/TIA.2026.3676945},
  publisher={IEEE}
}

@article{zhang2025hydrogen,
  title={Hydrogen energy storage system participated decentralized voltage control with multi-agent deep reinforcement learning method},
  author={Zhang, Xian and Gu, Changlei and Wang, Hong and Wang, Guibin and Xu, Yinliang and Sayed, Ahmed Rabee},
  journal={IEEE Transactions on Industry Applications},
  volume={61},
  number={2},
  pages={2578--2588},
  year={2025},
  publisher={IEEE}
}

@article{zhao2025system,
  title={System dynamics analysis of interaction behaviors and pricing mechanisms in grid-hydrogen-vehicle system},
  author={Zhao, Ruoxuan and Yang, Qiming and Li, Gengfeng and Li, Minghao and Ji, Chenlin and Liu, Dafu and Xu, Ziwen and Shi, Jiaju and Bie, Zhaohong},
  journal={Applied Energy},
  volume={391},
  pages={125909},
  year={2025}
}

@article{abdelghany2026optimal,
  title={Optimal configuration and coordinated scheduling for hydrogen-based mobility in green transit networks: dispatchable operations and storage control},
  author={Abdelghany, Muhammad Bakr and Al-Durra, Ahmed and Mohamed, Moataz and El Moursi, Mohamed Shawky and Gao, Fei},
  journal={IEEE Transactions on Industry Applications},
  year={2026},
  doi={10.1109/TIA.2026.3695468},
  publisher={IEEE}
}

@article{zheng2025safe,
  title={Safe and economical operation of electric-hydrogen refueling stations: A temperature-aware scheduling approach under vehicle uncertainties},
  author={Zheng, Wendi and Chen, Rui and Shao, Zhenguo and Pan, Jianing and Zhu, Youzhe},
  journal={International Journal of Hydrogen Energy},
  volume={113},
  pages={249--260},
  year={2025},
  publisher={Elsevier}
}

@article{wu2023integrated,
  title={An integrated energy analysis framework for evaluating the application of hydrogen-based energy storage systems in achieving net zero energy buildings and cities in Canada},
  author={Wu, You and Zhong, Lexuan},
  journal={Energy Conversion and Management},
  volume={286},
  pages={117066},
  year={2023},
  publisher={Elsevier}
}

@article{li2024cooperative,
  title={Cooperative economic dispatch of EV-HV coupled electric-hydrogen integrated energy system considering V2G response and carbon trading},
  author={Li, Ruiqi and Ren, Hongbo and Wu, Qiong and Li, Qifen and Gao, Weijun},
  journal={Renewable energy},
  volume={227},
  pages={120488},
  year={2024},
  publisher={Elsevier}
}

@article{yao2025unified,
  title={A unified data-driven approach under deep reinforcement learning with direct control responses for microgrid operations},
  author={Yao, Fulong and Zhao, Wanqing and Forshaw, Matthew and Zhou, Wenju},
  journal={Knowledge-Based Systems},
  volume={325},
  pages={113844},
  year={2025},
  publisher={Elsevier}
}

@inproceedings{fujimoto2018addressing,
  title={Addressing function approximation error in actor-critic methods},
  author={Fujimoto, Scott and Hoof, Herke and Meger, David},
  booktitle={International conference on machine learning},
  pages={1587--1596},
  year={2018},
  organization={PMLR}
}

@article{knosala2021hybrid,
  title={Hybrid hydrogen home storage for decentralized energy autonomy},
  author={Knosala, Kevin and Kotzur, Leander and R{\"o}ben, Fritz TC and Stenzel, Peter and Blum, Ludger and Robinius, Martin and Stolten, Detlef},
  journal={international journal of hydrogen energy},
  volume={46},
  number={42},
  pages={21748--21763},
  year={2021},
  publisher={Elsevier}
}

@article{sawant2024dc,
  title={DC fast charging stations for electric vehicles: A review},
  author={Sawant, Vikram and Zambare, Pallavi},
  journal={Energy conversion and economics},
  volume={5},
  number={1},
  pages={54--71},
  year={2024},
  publisher={Wiley Online Library}
}

@article{atabay2024design,
  title={Design and techno-economic analysis of solar energy based on-site hydrogen refueling station},
  author={Atabay, Reyhan and Devrim, Y{\i}lser},
  journal={International Journal of Hydrogen Energy},
  volume={80},
  pages={151--160},
  year={2024},
  publisher={Elsevier}
}

@article{parra2019review,
  title={A review on the role, cost and value of hydrogen energy systems for deep decarbonisation},
  author={Parra, David and Valverde, Luis and Pino, F Javier and Patel, Martin K},
  journal={Renewable and Sustainable Energy Reviews},
  volume={101},
  pages={279--294},
  year={2019},
  publisher={Elsevier}
}

@misc{elexon_mip,
  author = {{Elexon}},
  title  = {Market Index Prices},
  year   = {2026},
  note   = {Available at: \url{https://bmrs.elexon.co.uk/market-index-prices}. Accessed: 2026-06-18}
}

@techreport{uk_bill_components,
  author       = {{UK House of Commons Library}},
  title        = {Domestic Energy Prices},
  institution  = {House of Commons Library},
  year         = {2025},
  note         = {Available at: \url{https://commonslibrary.parliament.uk/research-briefings/cbp-9491/}. Accessed: 2026-06-18}
}

@dataset{era5_timeseries,
  author    = {{Copernicus Climate Change Service (C3S)}},
  title     = {ERA5 Hourly Time-Series Data on Single Levels from 1940 to Present},
  year      = {2025},
  publisher = {Climate Data Store (CDS)},
  doi       = {10.24381/cds.adbb2d47}
}

@misc{eho_hydrogen_cost_2024,
  author = {{European Hydrogen Observatory}},
  title  = {Cost of Hydrogen Production},
  year   = {2024},
  note   = {Available at: \url{https://observatory.clean-hydrogen.europa.eu/hydrogen-landscape/production-trade-and-cost/cost-hydrogen-production}. Accessed: 2026-06-18}
}

@misc{traffic_poisson_arrivals,
  author = {{IIT Bombay}},
  title  = {Vehicle Arrival Models: Count},
  howpublished = {NPTEL Course: Traffic Flow Modelling},
  note   = {Available at: \url{https://www.civil.iitb.ac.in/tvm/nptel/532_ArrModel/web/web.html}. Accessed: 2026-06-18}
}

@article{kurtz2020predicting,
  title={Predicting demand for hydrogen station fueling},
  author={Kurtz, Jennifer and Bradley, Thomas and Winkler, Erin and Gearhart, Chris},
  journal={International Journal of Hydrogen Energy},
  volume={45},
  number={56},
  pages={32298--32310},
  year={2020},
  publisher={Elsevier}
}

@misc{ehts_dashboard,
  author       = {Fulong Yao},
  title        = {EHTS Dashboard: Interactive Visualization of the Electric--Hydrogen Transportation System},
  year         = {2026},
  note         = {Available at: \url{https://flyao123.github.io/EHTS-Dashboard/}. Accessed: 2026-06-18}
}

@article{wan2025coordinated,
  title={Coordinated operation of alternative fuel vehicle-integrated microgrid in a coupled power-transportation network: a Stackelberg--Nash game framework},
  author={Wan, Yuyang and Wang, Ning and Du, Ershun and Liu, Xueshan and Wang, Yanbo and Chen, Zhe and Kang, Chongqing},
  journal={Applied Energy},
  volume={401},
  pages={126800},
  year={2025},
  publisher={Elsevier}
}

@article{fang2023optimal,
  title={Optimal energy management of multiple electricity-hydrogen integrated charging stations},
  author={Fang, Xiaolun and Wang, Yubin and Dong, Wei and Yang, Qiang and Sun, Siyang},
  journal={Energy},
  volume={262},
  pages={125624},
  year={2023},
  publisher={Elsevier}
}

@article{zhou2024comprehensive,
  title={A comprehensive survey of low-carbon planning and operation of electricity, hydrogen fuel, and transportation networks},
  author={Zhou, Yeao and Chen, Sheng and Chen, Jiayu},
  journal={IET Energy Systems Integration},
  volume={6},
  number={2},
  pages={89--103},
  year={2024},
  publisher={Wiley Online Library}
}

@article{zheng2014queuing,
  title={Queuing-based approach for optimal dispenser allocation to hydrogen refueling stations},
  author={Zheng, Jinyang and Zhao, Lei and Ou, Kesheng and Guo, Jinxing and Xu, Ping and Zhao, Yongzhi and Zhang, Lin},
  journal={International journal of hydrogen energy},
  volume={39},
  number={15},
  pages={8055--8062},
  year={2014},
  publisher={Elsevier}
}

@article{brown2022analysis,
  title={Analysis of customer queuing at hydrogen stations},
  author={Brown, Tim and Kisting, Hilary},
  journal={International Journal of Hydrogen Energy},
  volume={47},
  number={39},
  pages={17107--17120},
  year={2022},
  publisher={Elsevier}
}

@article{yang2018data,
  title={Data center holistic demand response algorithm to smooth microgrid tie-line power fluctuation},
  author={Yang, Ting and Zhao, Yingjie and Pen, Haibo and Wang, Zhaoxia},
  journal={Applied Energy},
  volume={231},
  pages={277--287},
  year={2018},
  publisher={Elsevier}
}

@article{schulman2017proximal,
  title={Proximal policy optimization algorithms},
  author={Schulman, John and Wolski, Filip and Dhariwal, Prafulla and Radford, Alec and Klimov, Oleg},
  journal={arXiv preprint arXiv:1707.06347},
  year={2017}
}

\end{document}